\documentclass[a4paper,10pt]{article}
\usepackage{science_vielva}

\usepackage{amsmath}
\usepackage{graphicx}
\usepackage{graphics}
\usepackage{amssymb}

\newcommand{\nside}{\textsc{\small{N\tiny{SIDE}}} }
\newcommand{\nsidep}{\textsc{\small{N\tiny{SIDE}}} }
\newcommand{\cs}{Cold Spot }
\newcommand{\mas}{\mathrm{MAX} }
\newcommand{\masr}{\mathrm{MAX_R} }
\newcommand{\csp}{Cold Spot}
\newcommand{\ii}{{\'\i}}
\newcommand{\wcr}{{}w\left(\theta,\phi;R\right)}
\newcommand{\wcri}{{}w\left(\theta_i,\phi_i;R\right)}
\newcommand{\hwcri}{{}\hat{w}\left(\theta_i,\phi_i;R\right)}
\newcommand{\npix}{\mathrm{N_{pix}}}

\title{A comprehensive overview of the \cs}
\author{Patricio Vielva}
\date{\small{Instituto de F\ii sica de Cantabria (CSIC - Univ. de Cantabria) \\ 
Avda. Los Castros s/n, E39005 - Santander, Spain\\
e-mail : vielva@ifca.unican.es}}

\begin{document}
\maketitle

\begin{abstract}
The report of a significant deviation of the CMB temperature anisotropies distribution from Gaussianity (soon after the public release of the WMAP data in 2003) has become one of the most solid WMAP \emph{anomalies}. This detection grounds on an excess of the kurtosis of the Spherical Mexican Hat Wavelet coefficients at scales of around 10 degrees. At these scales, a prominent feature |located in the southern Galactic hemisphere| was highlighted from the rest of the SMHW coefficients: the \csp.
This article presents a comprehensive overview related to the study of the \csp, paying attention to the non-Gaussianity detection methods, the morphological characteristics of the \csp, and the possible sources studied in the literature to explain its nature. Special emphasis is made on the \cs compatibility with a \emph{cosmic texture}, commenting on future tests that would help to give support or discard this hypothesis.
\end{abstract}

\begin{body}

\section{Introduction}
\label{intro}
Besides the great success of the NASA WMAP satellite on providing a detailed knowledge of the cosmological parameters that define the physical properties of the Universe (e.g.,~\cite{komatsu09, larson10}), some unexpected results have attracted the attention of the cosmological community soon after the first release of the
WMAP data: the so-called \emph{WMAP anomalies}. Some of these anomalies are related to hemispherical asymmetries (e.g.,~\cite{eriksen04a,hansen04a,eriksen04b,hansen04b,donoghue05,hoftuft09,pietrobon10,vielva10a,paci10}), an anomalous alignment of the quadrupole and octopole components (e.g.,~\cite{bielewicz04,schwarz04,copi04,deOliveira04,bielewicz05,land05,copi06,abramo06,gruppuso10,frommert10}), significantly low variance of the CMB temperature fluctuations (e.g.,~\cite{monteserin08,ayaita10,cayon10,cruz10}), or anomalous alignment of the CMB features toward the Ecliptic poles (e.g.,~\cite{wiaux06,vielva07a}). Some of these aspects are addressed in this special issue.
In addition to the previous findings, the prominent cold spot (hereinafter, the \csp) detected in the southern hemisphere by~\cite{vielva04} became one of the most studied anomalies of the WMAP data.
The \cs was identified after testing that the Spherical Mexican Hat Wavelet (SMHW) coefficients of the WMAP data presented an excess of kurtosis (at scales of around $10^\circ$ in the sky), as compared to the distribution derived from isotropic and Gaussian CMB simulations.
This paper presents a complete review on the detection and characterization of this non-standard signature, and a description of the different attempts made so far in understanding what could be the cause behind such departure from the standard inflationary paradigm. It is organized as follows: in Section~\ref{sec:ng_detect}, I justify the use of wavelets as a natural tool for probing the Gaussianity of the CMB temperature fluctuations. I also present the different statistics applied in the wavelet space that had led to point out the WMAP data incompatibility with the standard model.
In section~\ref{sec:chare}, I briefly describe the morphological characteristics of the \csp. The important question of the actual significance of the detection of the \csp, and the aspects associated with \emph{a posteriori} interpretations are addressed in Section~\ref{sec:sig}. 
Some of the different sources that have been considered in the literature to explain the \cs feature are discussed in Section~\ref{sec:sources}. 
In Section~\ref{sec:texture}, I explain in detail a plausible hypothesis to accommodate the existence of the \cs together with the standard cosmological model: a cosmic texture. In addition, I also describe possible follow-up tests that could help to confirm or discard such hypothesis. Finally, my conclusions are given in Section~\ref{sec:final}.

\section{The non-Gaussianity detection}
\label{sec:ng_detect}
The \cs was firstly identified through a \emph{blind} Gaussianity test of the WMAP first year data~\cite{vielva04}. This test was designed to probe the isotropic and Gaussian nature of the CMB, as predicted by the standard inflationary model (see, for instance~\cite{liddle00}), and it was based in a multiresolution analysis performed with the Spherical Mexican Hat Wavelet (SMHW). In this Section, I summarize the non-Gaussianity detection that led to the identification of the \csp. I start by justifying why an analysis based on wavelets was proposed, and presenting the main characteristics of the wavelet used in the analysis: the SMHW. Afterwards, I explain which statistics (all of them based on the SMHW coefficients) reported the original deviations from Gaussianity of the WMAP data.

\subsection{Why a wavelet?}
\label{subsec:wave}
Nowadays, the CMB scientific community is already very familiar with the application of wavelets (and other members of the \emph{-lets} zoo, like curvelets, ridglets or needlets) to data analysis. However, it is worth recalling that this is a relatively new custom. Although wavelet applications in cosmology shyly started already in mid 80s, it was not until 1997 that the first application to CMB was presented~\cite{ferreira97}, precisely in a exercise devoted to probe the Gaussianity of the CMB; and it was just a year afterwards that the first CMB data analyses with wavelets came to light~\cite{popa98,pando98}, in particular, with two applications to COBE~\cite{bennett96} data.
During the last decade, the application of wavelets to extract cosmological information from CMB data has growth considerably, in many different branches: compact source detection (e.g.~\cite{tenorio99,cayon00,vielva03,gonzaleznuevo06,lopezcaniego06,pires06}), 
Gaussianity (e.g.~\cite{aghanim99,mukherjee00,barreiro00,barreiro01,cayon01,martinezgonzalez02,vielva04,cruz05,mcewen05,pietrobon08,wiaux08,curto09,rudjord09}),  cross-correlation with large scale structure (e.g.~\cite{vielva06,pietrobon06,liu06,mcewen08}), decomposition of the coupled E/B signals~\cite{cao09}, probing isotropy (e.g.~\cite{wiaux06,vielva07a,pietrobon10}), cosmic string detection~\cite{hammond09}, microwave sky recovery (e.g.~\cite{maisinger04,moudden05,hansen06,delabrouille09}), 
CMB denoising (e.g.~\cite{sanz99,tenorio99}), CMB power spectrum determination~\cite{fay08}, and primordial power spectrum recovery (e.g.~\cite{mukherjee03}) are some of the application fields.
I refer the reader to~\cite{mcewen07,vielva07b} for some reviews on the wavelets applications to CMB data, with a particular emphasis on data analysis on the sphere.

The wavelet transform (e.g.~\cite{odgen97,antoine04}) has become very popular for a major reason: they offer a unique opportunity to probe scale-dependent phenomena, but keeping, at the same time, information about spatial localization. This is a clear advantage |for many purposes| over classical Fourier or harmonic transforms: physical processes typically exhibit a clear scale-dependent behaviour, and, often, such behaviour differs enough from one phenomenon to another (e.g., in a microwave image, the localized emission due to cluster of galaxies has very different properties as compared to the large-scale signal produced by the Galactic components).

The capability of emphasising or amplifying some features (at a particular scale) makes wavelets unique to probe the Gaussianity of the CMB: there are many different physical processes that might introduce non-Gaussian signatures into a CMB signal, at a very particular scale range (e.g., primordial non-Gaussianity due to non-standard inflationary scenarios, cosmic defects like strings or textures, secondary anisotropies, foreground emissions, \ldots). 

%%%
\begin{figurehere}
\begin{center}
\includegraphics[width=9cm,keepaspectratio]{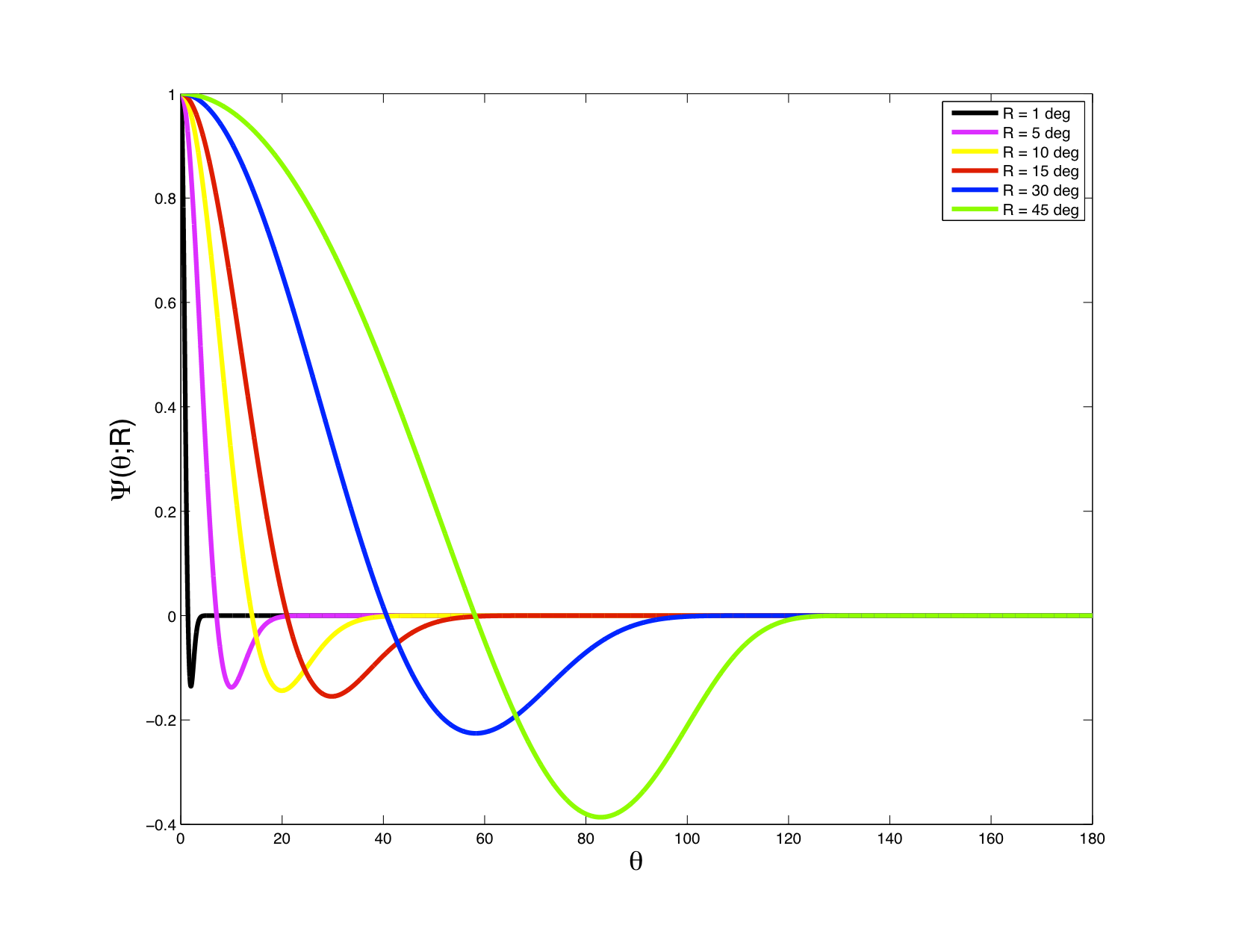}
\caption{~\label{fig:smhw_profile}Figure shows the profile on the SMHW as a function of the angular distance $\theta$ for different wavelet scales $R$, ranging from $1^\circ$ to $45^\circ$. Notice that the amplitude of the SMHW at $\theta \equiv 0$ has been fixed to 1, to allow for an easier comparison.}
\end{center}
\end{figurehere}
%%%%

In this sense, when a Gaussianity analysis of the CMB is performed in wavelet space, we are in a very adequate framework to probe, almost separately, any potential non-Gaussian signatures present in the data.
The ability of wavelets to amplify a given signature is explained because they can be seen as compensated filters (i.e., $\int_{\Re^n} \Psi = 0$, where $\Re^n$ is the space where the wavelet $\Psi$ is defined). This property is not satisfied by other standard filters, as the top-hat or the Gaussian functions. As mentioned above, wavelets do not only provide us with the capability of selecting a given scale range, but they also allow one to keep the information about spatial localization: we do not just study, for instance, the compatibility with Gaussianity, but we are also able to identify where in the data a deviation might be spatially localized. This intrinsic property of the wavelet transform (spatial localization) is also unique to explore not only Gaussianity, but also isotropy, since the statistical properties can be studied (almost independently) from one region in the data to another, in a self-consistent way.
Hereinafter I will focus in the use of the two-dimensional (2D) continuous wavelet transform (CWT, see, for instance, Chapter 2 in~\cite{antoine04}), and, in particular, on the 2D CWT defined on the sphere. There are different ways to define wavelets on the sphere (e.g.~\cite{antoine98,wiaux05,sanz06,wiaux07}); most of the works related to the \cs have been performed using the definition of the isotropic Spherical Mexican Hat Wavelet (SMHW) proposed by~\cite{martinezgonzalez02} which is a stereographic projection of the Mexican Hat Wavelet, as proposed by~\cite{antoine98}:
%%%
\begin{equation}
\label{eq:smhw}
\Psi\left(\theta;R\right) = \frac{1}{\sqrt{2\pi}RN_R}\left[1 + \left(\frac{y}{2}\right)^2\right]^2 \left[2-\left(\frac{y}{R}\right)^2\right]e^{-\frac{y^2}{2R^2}},
\end{equation}
%%%
where $y \equiv 2\tan\frac{\theta}{2}$ is the stereographic projection variable, $\theta \in \left[0, \pi\right)$ is the co-latitude, and the constant $N_R \equiv \sqrt{1 + \frac{R^2}{2} + \frac{R^4}{4}}$ is chosen such as the square of the wavelet function $\Psi\left(\theta;R\right)$ is normalized to unity.
In figure~\ref{fig:smhw_profile} the radial profile of the SMHW, for different wavelet scales $R$, is shown.
For a signal $T\left(\theta,\phi \right)$ defined on the sphere, its spherical harmonic coefficients $t_{\ell m}$ are defined as:
%%%
\begin{equation}
\label{eq:tlm}
t_{\ell m} = \int \mathrm{d}\Omega Y_{\ell m}^* \left(\theta, \phi \right)T\left(\theta,\phi \right),
\end{equation}
%%%
where $\mathrm{d}\Omega = \mathrm{d}\theta \sin \theta \mathrm{d}\phi$, the spherical coordinates are, as mentioned above, the co-latitude $\theta$  (related to the latitude $b$ as $b = \pi/2 - \theta$), and $\phi \in \left[0,2\pi\right)$ is the longitude. The function $Y_{\ell m} \left(\theta, \phi \right)$ is the spherical harmonic of order $\ell$ and $m$, and $·^*$ denotes complex conjugation.
For an isotropic 2D CWT on the sphere the wavelet coefficients $\wcr$ are obtained as:
%%%
\begin{equation}
\label{eq:wc}
w\left(\theta,\phi;R \right) = \sum_{\ell = 0}^{\ell_{\mathrm{max}}} \sum_{m = -\ell}^{\ell} t_{\ell m} \Psi_\ell \left(R\right) Y_{\ell m} \left(\theta, \phi \right)T\left(\theta,\phi \right),
\end{equation}
%%%
where $\Psi_\ell \left(R\right)$ is the window function associated with the wavelet function (e.g., $\Psi\left(\theta;R\right)$ in equation~\ref{eq:smhw}), and $\ell_{\mathrm{max}}$ represents the maximum multipole associated with a given resolution of the signal $T\left(\theta,\phi \right)$, typically limited by the size of the pixel adopted to represent such signal on the sphere.

\subsection{The statistics}
\label{subsec:statis}

The \cs was firstly detected via a positive deviation of the kurtosis of the SMHW coefficients at scales of around $R \approx 300^\circ$. The inspection of the
map of the SMHW coefficients at these scales revealed the presence of a very large and cold spot in the southern hemisphere. The comparison of the amplitude and the area of this cold spot as compared with Gaussian simulations showed that it was particularly anomalous. Finally, a higher criticism test of the SMHW coefficients also indicated a deviation, at around the same wavelet scales, also showing that the major source for such deviation was located in the position of the \csp. In the following subsections I describe briefly these statistics: the kurtosis, the amplitude, the area and the higher criticism.

%%%
\begin{figurehere}
\begin{center}
\includegraphics[width=8cm,keepaspectratio]{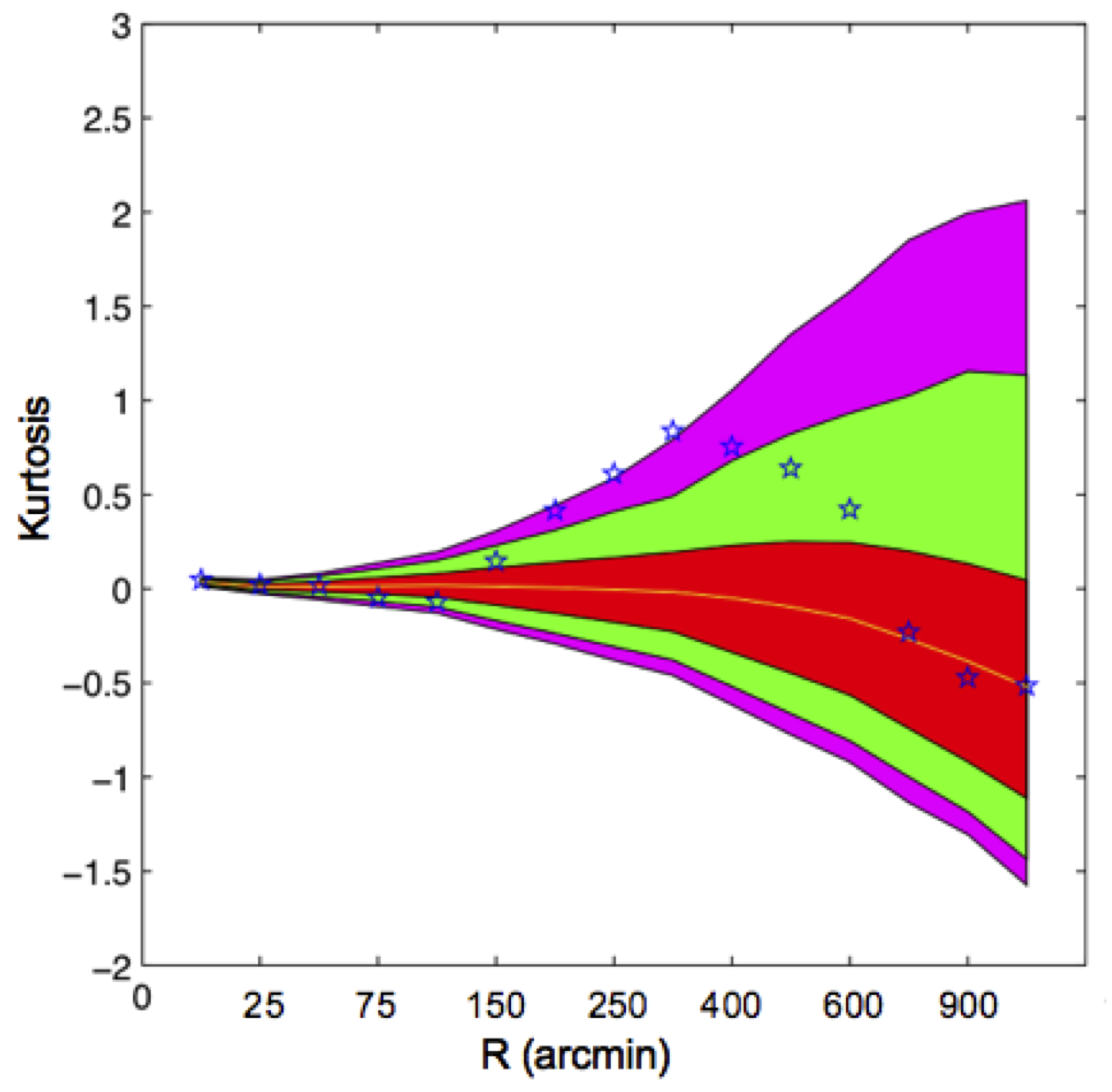}
\caption{~\label{fig:kurtosis}Figure showing the positive deviation of the kurtosis of the wavelet coefficients at scales $R$ of 250 and 300 arcmin found by~\cite{vielva04}. Red, green and magenta regions represent the acceptance intervals at 32\%, 5\% and 1\%, respectively.}
\end{center}
\end{figurehere}
%%%%
\subsubsection{The kurtosis}
\label{subsubsec:kurto}
%%%
\begin{figure*}
\begin{center}
\includegraphics[width=8cm,keepaspectratio]{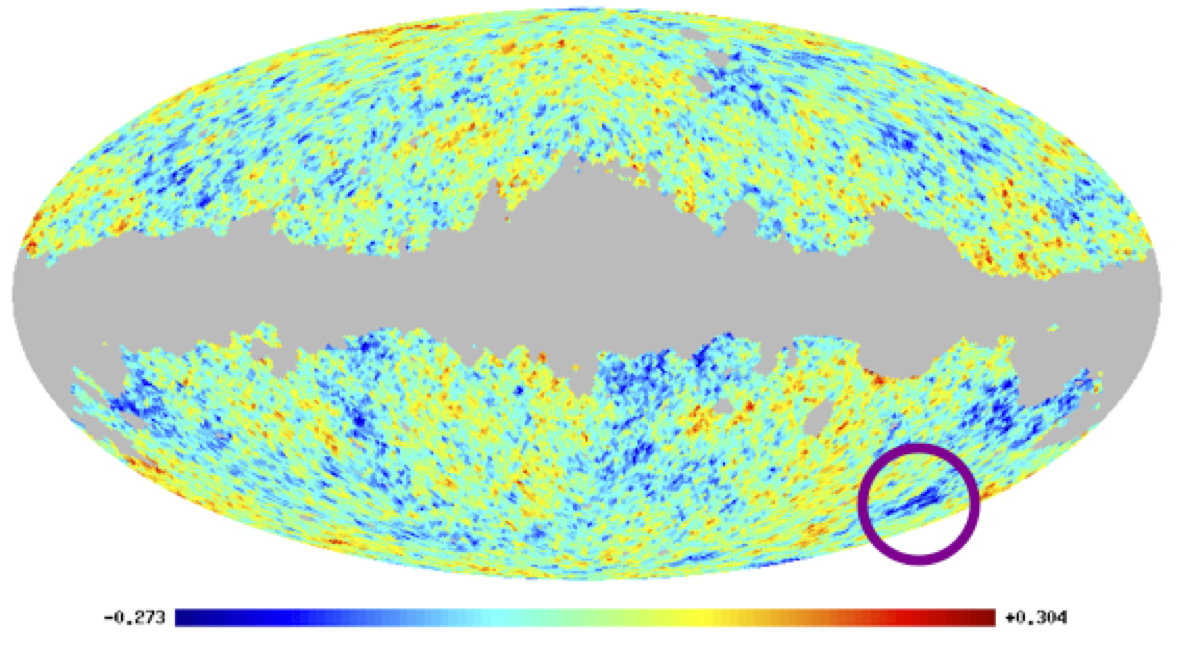}
\includegraphics[width=8cm,keepaspectratio]{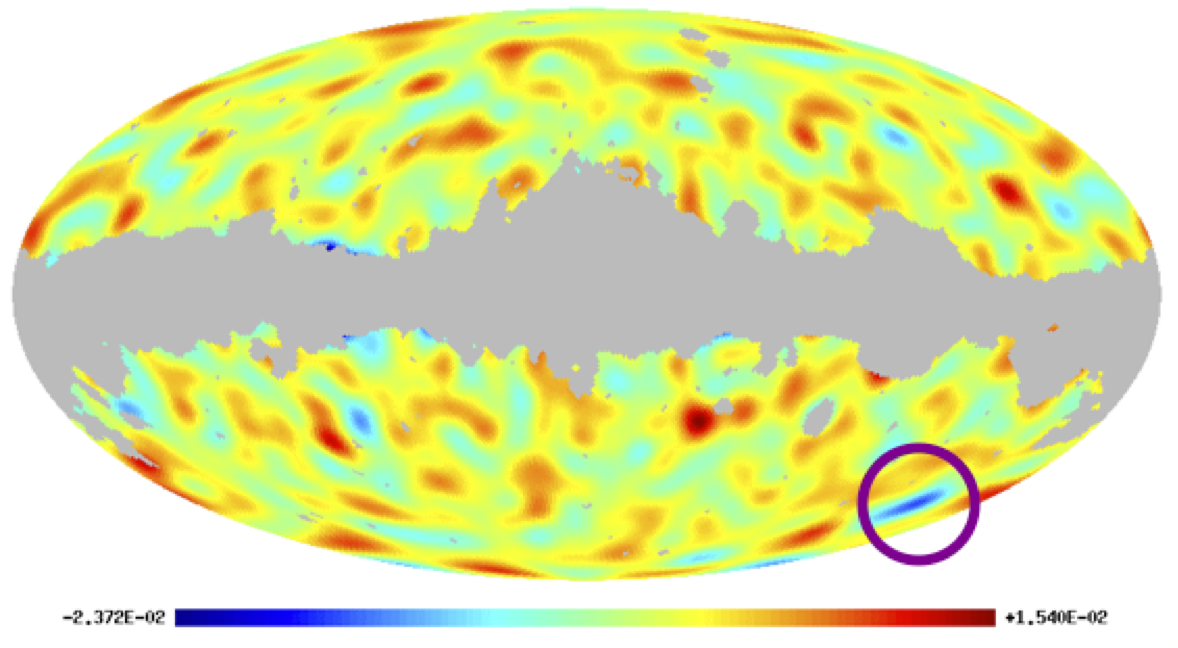}
\caption{~\label{fig:real_smhw}Left panel: CMB cleaned map derived from the 4th release of the WMAP data, obtained via the template fitting technique described in~\cite{gold09}. Right panel: wavelet coefficients of the previous map, obtained after the SMHW convolution at a scale of $R=250$ arcmin. The location of the \cs is indicated in both panels by the circle. The centre of the \cs is $\theta = 147^\circ$ and $\phi = 210^\circ$.}
\end{center}
\end{figure*}
%%%%
The two most obvious indicators for a possible deviation from Gaussianity of a given data set $z=\{z_1,z_2,\ldots,z_N\}$ are the third ($\mu_3$) and fourth ($\mu_4$) central moments. The $n$-central moment of a random set of $N$ numbers is defined as:
%%%
\begin{equation}
\label{eq:cm}
\mu_n = \frac{1}{N}\sum_{i=1}^N \left({z_i - \bar{z}}\right)^n,
\end{equation}
%%%
where $\bar{z} = (1/N)\sum_{i=1}^N z_i$ is the mean value of the data set.
For a Gaussian random data set of zero mean (i.e., $\mu_1 = \bar{z} \equiv 0$) and dispersion $\sigma$ (i.e., $\mu_2 \equiv \sigma^2$), it is trivial to prove that higher order central moments are either zero (in the case of $\mu_{2n+1}$, for $n \ge 1$) or a given function of the dispersion (in the case of $\mu_{2n}$, for $n \ge 1$). 
Usually, it is much more convenient to work with normalized central moments $\nu_n$, such as:
%%%
\begin{equation}
\label{eq:ncm}
\nu_n = \frac{\mu_n}{\sigma^n}. 
\end{equation}
%%%
The normalized central moments are more convenient than central moments, since they are referred to the intrinsic fluctuations of the random data set (represented by the dispersion). This helps to absorb into $\nu_n$ possible uncertainties on the knowledge on the amplitude of the fluctuations of the random sample. The normalized central moment $\nu_3$ is normally referred to as the skewness ($S$), whereas $K\equiv\nu_4 - 3$ is called kurtosis.
The reason for the subtraction of the number $3$ in the previous expression comes from the fact that, for a Gaussian random variable, $\mu_4 = 3\sigma^4$ and, therefore, $\nu_4 = 3$. The previous definition assures that, as it happens for the skewness ($S$), the kurtosis ($K$) of a Gaussian field is zero.
Applying these concepts to the SMHW coefficients $\wcr$, the skewness $S_R$ and the kurtosis $K_R$ of the wavelet coefficients, as a function of the wavelet scale $R$, can be defined as follows:
%%%
\begin{eqnarray}
S_R & = & \frac{1}{\sigma^3_R}\frac{1}{\npix(R)}\sum_{i=1}^{\npix(R)}\wcri^3\label{eq:skew} \\
K_R & = & \frac{1}{\sigma^4_R}\frac{1}{\npix(R)}\sum_{i=1}^{\npix(R)}\wcri^4 - 3,\nonumber\label{eq:kurto}
\end{eqnarray}
%%%
where, at each scale, it is assumed that the coefficients $\wcr$ have zero mean. $\sigma_R$ is the dispersion of the wavelet coefficients at the scale $R$:
%%%
\begin{eqnarray}
\sigma_R & = & \left[ \frac{1}{\npix(R)}\sum_{i=1}^{\npix(R)}\wcri^2\label{eq:disp}\right]^{1/2}.
\end{eqnarray}
%%%
In the previous expressions, $\npix(R)$ represents the number of wavelet coefficients at a given scale $R$.
Notice that, very often, CMB data is not available in the full celestial sphere (for instance because strong contamination from astrophysical foregrounds has to be masked).
In most of the works in the CMB field, the HEALPix tessellation~\cite{gorski05} is adopted. In this scheme, the resolution of a given image represented on the sphere is given by the \nside parameter, which indicates how many divisions of the 12 basic pixels are required to achieve such resolution. The $\nside$ parameter is related with the number of the pixels ($\npix$) required to fill the sphere at that resolution as:
$\npix = 12$\nsidep$^2$.
In figure~\ref{fig:kurtosis} the major result of the seminal work \cite{vielva04} on the \cs is shown. It represents the kurtosis of the wavelet coefficients, as a function of the scale, for the first release of the WMAP data (blue stars). The solid yellow line represents the mean value obtained from 10,000 CMB Gaussian random simulations, taking into account the instrumental properties of the analyzed data, and generated from an angular power spectrum derived form the best-fit cosmological model. The coloured regions (red, green and magenta) represent the acceptance intervals at 32\%, 5\% and 1\%, respectively. Notice that at $R$ scales of 250 and 300 arcmin, the kurtosis of the WMAP data was above the 1\% acceptance interval.
In detail, the excess of kurtosis is given by a p-value of $\approx 4\times10^{-3}$ for the two scales. No significant deviations were found related to the skewness of the wavelet coefficients.
%%%
\begin{figure*}
\begin{center}
\includegraphics[width=8cm,keepaspectratio]{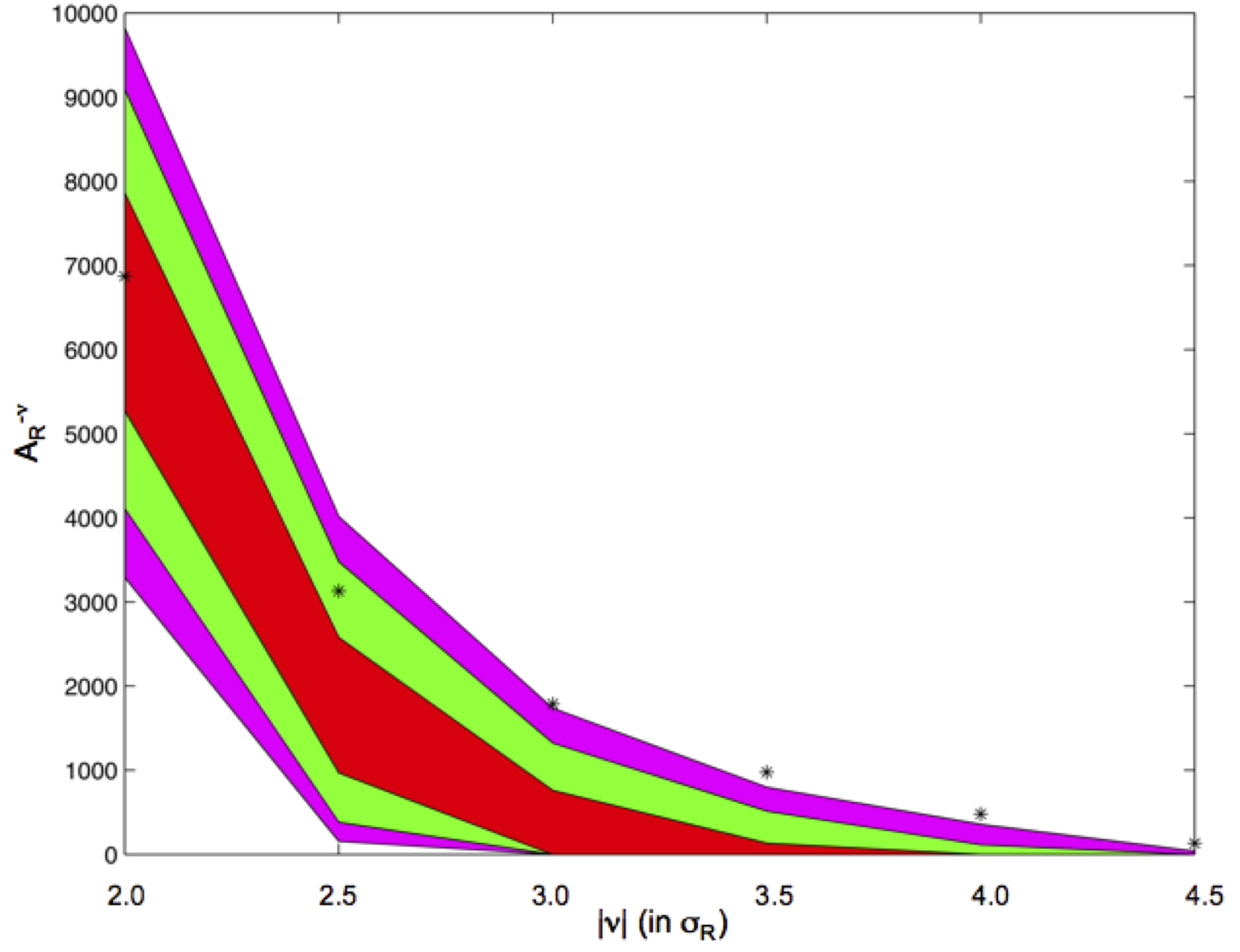}
\includegraphics[width=8cm,keepaspectratio]{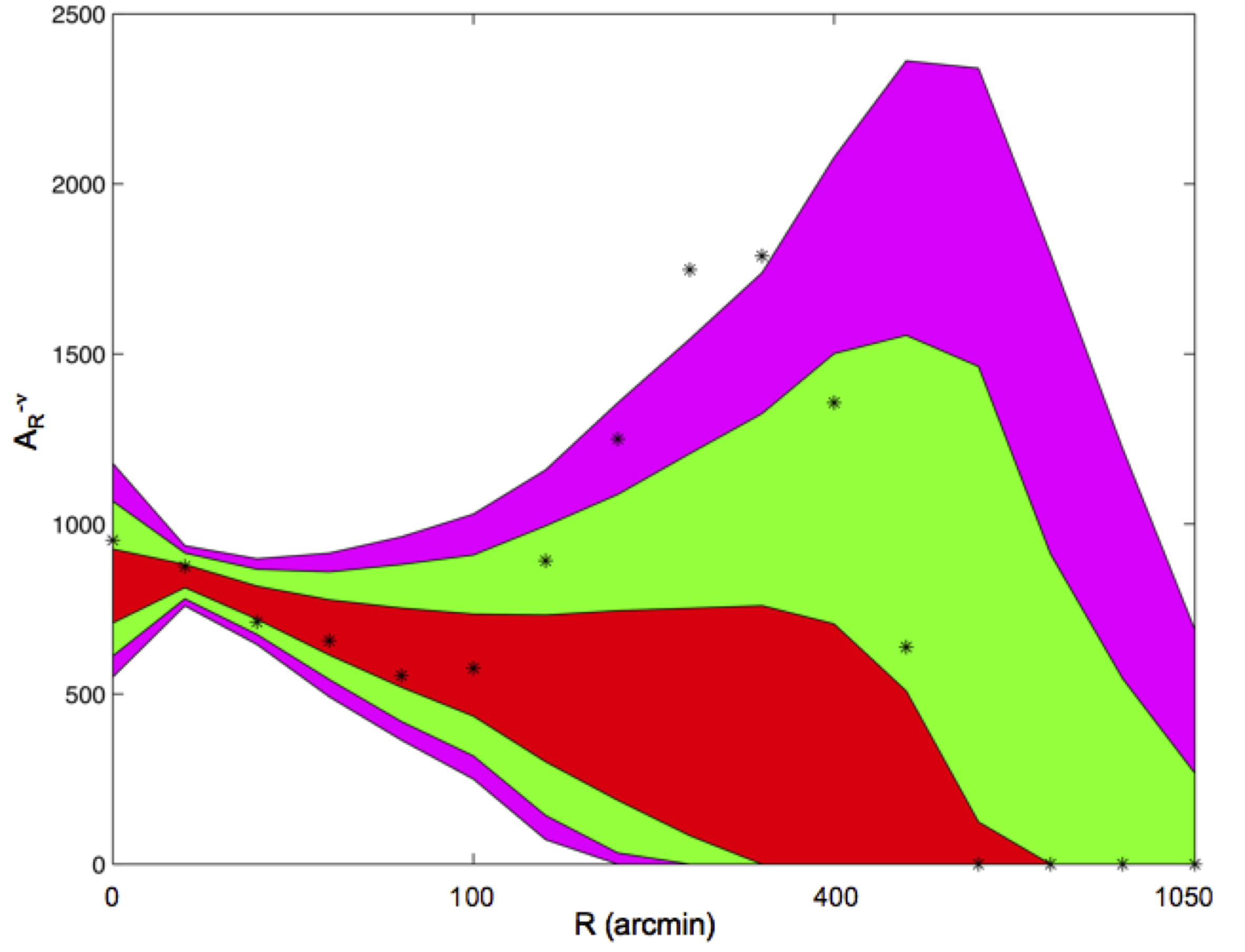}
\caption{~\label{fig:cold_area}Right panel: cold area of the SMHW coefficients (at $R=300$ arcmin), as a function of the threshold ($\nu$).
Left panel: cold area ($A^{-\nu}_R$) of the SMHW coefficients (at $\nu = 3\sigma_R$), as a function of the scale ($R$). 
As in figure~\ref{fig:kurtosis}, the red, green, and magenta regions represent the 32\%, 5\% and 1\% acceptance intervals, respectively.
These plots correspond to the analysis done by~\cite{cruz05}.}
\end{center}
\end{figure*}
%%%%
The analysis was repeated in the two Galactic hemispheres separately. This was motivated by previous findings of asymmetries related to the genus~\cite{park04} and the N-point correlation function~\cite{eriksen04a}. It was found that whereas there was not deviation on the northern hemisphere, the excess of kurtosis was even more remarkable in the southern region. In particular, at the scale of 250 arcmin, the kurtosis of the SMHW coefficients was associated with a p-value of $\approx 2\times10^{-3}$. Again, no deviation on the skewness was noticed. Hence, these analyses indicated that the source for the deviation on the kurtosis of the wavelet coefficient was related to feature/s with a typical length of around $10^\circ$ in the sky, and located in the southern hemisphere.
Thanks to the frequency/spatial properties of the SMHW decomposition, it was possible to study in detail such features. In the left panel of figure~\ref{fig:real_smhw}, the cleaned CMB map obtained from the fourth release of the WMAP data is given. In particular, I plot the optimal combination for cosmological analysis (hereinafter, the
QVW map), obtained as a noise-weighted linear combination of the Q-, V-, and W-bands, previously cleaned via a template fitting (e.g.~\cite{gold09}). Regions highly contaminated by Galactic foregrounds, nearby clusters, and extragalactic point sources, have been masked. In the right panel, the SMHW coefficients of the previous map are represented, at the scale $R=250$ arcmin. It is evident the presence of the \cs in the southwest side of the image.
In particular, the centre of the \cs is estimated to be $\theta = 147^\circ$ and $\phi = 209^\circ$. 
The study of the \csp, through the application of follow-up tests, provided further evidences for its anomalous nature. Let me review these tests in the following subsections.

\subsubsection{The amplitude}
\label{subsubsec:ampli}
One of the most trivial statistics to study extreme values (as the \csp) in a random sample is the largest/smallest observation. In~\cite{vielva04} it was established that the temperature of the \cs was -4.57 times the dispersion of the SMHW coefficients at $R=250$ arcmin. This cold value represented a p-value of 0.01 (relative to Monte Carlo simulations). A more robust statistic related to the extreme values is the $\mas$ statistic, understood as the largest observation (in absolute value). For the particular case of the SMHW coefficients, $\mas$ is defined, at scale $R$, as:
%%%
\begin{equation}
\label{eq:max}
\masr = \max\left\{\left| \wcri \right|\right\}. 
\end{equation}
%%%
The $\mas$ statistic is more robust than selecting the coldest of the extrema, since the selection of the lowest values could be seen as an \emph{a posteriori selection}. This statistic was studied in~\cite{cruz07a}, showing that the \cs was always the maximum absolute observation of the WMAP data at scales around 300 arcmin, representing an upper tail probability of 0.38\% (relative to Monte Carlo simulations). This value was less significant than the one mentioned in the previous subsection. The reason for this change is, as commented, that the $\mas$ statistic is more robust than simply selecting the smallest values of the observations.

\subsubsection{The area}
\label{subsubsec:area}
The area above or below a given threshold is one of the most common statistics used to characterize the properties of a random field. In particular, the area is the most commonly Minkowski functional used in the literature\footnote{For 2D images, there are three Minkowski functionals, namely the contour or length, the area, and the genus. These three quantities are defined above/below a given threshold.} (see, for instance,~\cite{coles88,gott90,schmalzing98}). Generalizing this concept to the case of the wavelet coefficients, we can define cold ($A^{-\nu}_R$) and hot ($A^{+\nu}_R$) areas, at a given threshold $\nu$ and a given scale $R$, as:
%%%
\begin{eqnarray}
\label{eq:areas}
A^{-\nu}_R & = & \#\left\{ \wcri < -\nu \right\}, \\
A^{+\nu}_R & = & \#\left\{ \wcri \geq +\nu \right\}, \nonumber
\end{eqnarray}
%%%
where the \emph{number operator} $\#\{\mathtt{condition}_i\}$ indicates how many times $\mathtt{condition}_i$ is satisfied, for $i$ ranging from 1 to $\npix(R)$. 
The cold and hot areas of the WMAP data were analyzed by~\cite{cruz05}. It was reported that, whereas the hot area was consistent with the expected behaviour for the standard Gaussian model (at all the scales $R$ and thresholds $\nu$), the cold area was not compatible. In particular, deviations from Gaussianity were found, again, at SMHW scales of $R\approx 300^\circ$. The deviation took place for thresholds equal or smaller than $-3\sigma_R$ (see figure~\ref{fig:cold_area}). The analysis per different regions of the sky confirmed that the anomaly on the cold area was localized in the southern-west Galactic quadrant of the sky, and that the \cs was responsible for this anomaly. In particular, the cold area of the WMAP data (at the mentioned scales, and below a threshold of $-3\sigma_R$) was found anomalous with a probability of $\approx 99.7\%$, whereas it became fully compatible with
Gaussian simulations, once the \cs was not considered in the analysis.
As for the case of the $\mas$ statistic, a more conservative estimator (i.e., less dependent from the fact that the \cs is negative) can be considered, just by selecting the maximum value of the previous cold and hot areas:
%%%
\begin{equation}
\label{eq:max}
A^{\nu}_R = \max\left\{A^{-\nu}_R, A^{+\nu}_R\right\}. 
\end{equation}
%%%
This new statistic was used by~\cite{cruz07a}, finding, again, that the WMAP data was anomalous about thresholds larger than $|3\sigma_R|$, for scales of the SMHW of around 300 arcmin.

\subsubsection{The Higher criticism}
\label{subsubsec:hg}
Higher criticism (HC) is a relatively new statistic introduced in 2004 by~\cite{donoho04}, and firstly applied to the context of probing the Gaussianity of the CMB only a year after by~\cite{cayon05}. Although there is not a unique definition for the HC, all the forms proposed in the literature satisfy the same key concept: HC is a measurement of the distance between a given sample of $n$ elements to a Gaussian probability density distribution, established by means of the difference between the p-value $p_i$ of a given observation $X_i$ |assuming it comes from a N$(0,1)$|, and its cardinal position on the sorted list (in increasing order) of p-values $p_i$ (i.e., $p_{i-1} < p_i < p_{i+1}, \forall~ i=1,\ldots,n$). The HC associated with the sample is just defined as the largest value of such differences. 

This concept can be applied to the SMHW coefficients of a given signal (e.g., the QVW map) at a given scale $R$. This was the analysis proposed by~\cite{cayon05}. Let us adopt the following definition for the HC associated with $\npix(R)$ wavelet coefficients $\wcri$, at scale $R$:
%%%
\begin{equation}
\label{eq:hc}
\mathrm{HC_\npix (R)} = \max\left\{ HC^i_{\npix} (R) \right\},
\end{equation}
%%%
where the maximization is made over the quantity $HC^i_{\npix} (R)$, that provides the difference between the \emph{experimental} probability of the wavelet coefficients $\wcri$ at scale $R$ and the corresponding \emph{theoretical} p-value. Such quantity reads as:
%%%
\begin{equation}
\label{eq:hci}
HC^i_{\npix} (R) = \sqrt{\npix}\frac{\left\vert \frac{i}{\npix(R)} - p_i(R) \right\vert}{\sqrt{p_i(R)\left(1-p_i(R)\right)}},
\end{equation}
%%%
where the p-value is given by $p_i(R) = P\{ |\mathrm{N}(0,1)| > |\hwcri| \}$. The $\hat{}$ operator indicates that the $\npix(R)$ SMHW coefficients at the scale $R$ have been transformed into a zero mean and unit variance sample. Let me remark that the p-values have been sorted in increasing order.
%%%
\begin{figurehere}
\begin{center}
\includegraphics[width=8cm,keepaspectratio]{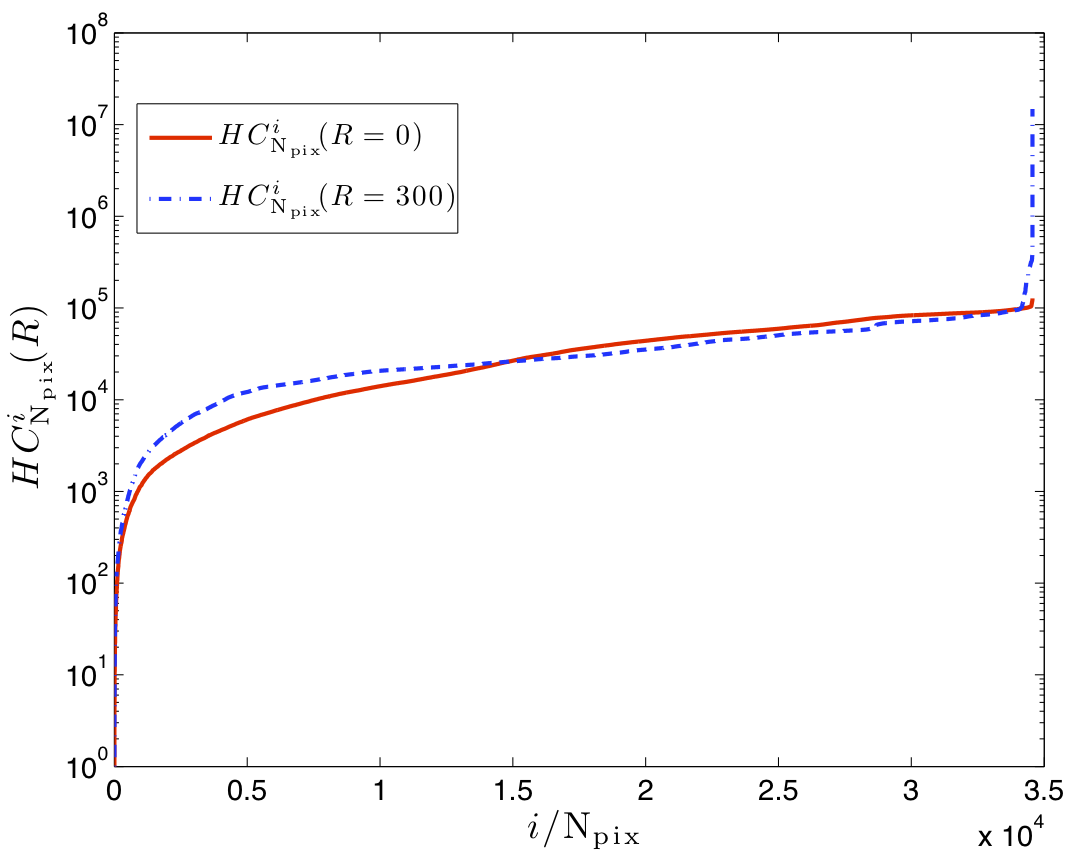}
\caption{~\label{fig:hcis}The $HC^i_{\npix}$ values obtained from the analysis of the QVW map. The solid red line corresponds to the application of the equation~\ref{eq:hci} to the CMB map in the real space (i.e., $R\equiv 0$), whereas the dot-dash blue line corresponds to the analysis of the SMHW coefficients at scale $R=300$ arcmin. $HC^i_{\npix}$ curves are normalized by their respective minimum values.}
\end{center}
\end{figurehere}
%%%%
%%%
\begin{figure*}
\begin{center}
\includegraphics[width=8cm,keepaspectratio]{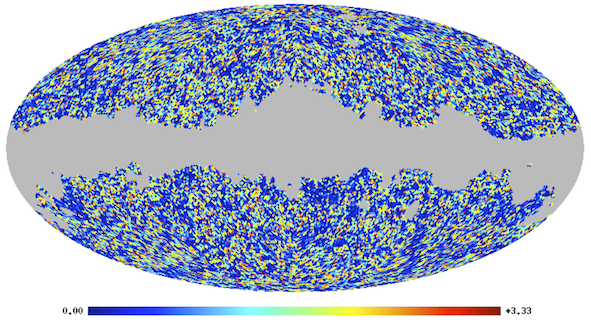}
\includegraphics[width=8cm,keepaspectratio]{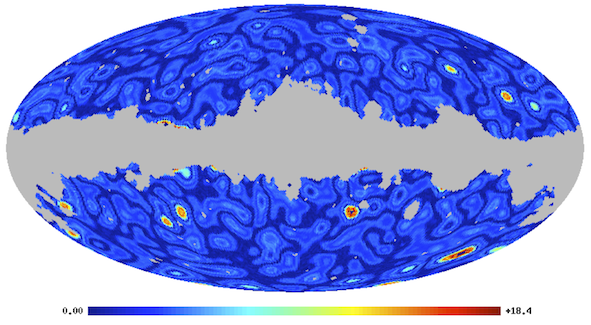}
\caption{~\label{fig:hci}Maps of $HC^i_{\npix}$ obtained from the analysis of the QVW map derived from the WMAP data. Left panel corresponds to the study of the real space case, while right panel shows the outcome of the analysis of the SMHW coefficients at $R=300$ arcmin. Whereas for the former there are not particular signatures in the map, the wavelet analysis shows some prominent features, being the \cs the most pronounced one.}
\end{center}
\end{figure*}
%%%% 

As explained in~\cite{cayon05}, the HC represents, under certain conditions, some advantages with respect to more traditional statistics designed to study the Gaussianity of a given sample. In particular, HC seems to be a better estimator than the $\mas$ statistic: whereas the latter is designed to capture Gaussian deviations caused by very large values of the distribution, the HC is also sensitive to anomalies produced by moderate values. In addition, the HC can identify which values (in a given sample) are the ones that differ from the theoretical Gaussian distribution.

In~\cite{cayon05} it was reported that the $\mathrm{HC_\npix (R)}$ was above the 1\% acceptance interval, again, at the SMHW scale $R=300$ arcmin. 
They found that, actually, all the SMHW coefficients associated with $HC^i_{\npix} (R)$ values above the 1\% acceptance interval set by CMB Gaussian simulations, where localized in the position of the \csp. This extra test was an additional support to the anomalous nature of the WMAP data, and of the \cs in particular. Results were confirmed by~\cite{cruz07a} for the analysis of the second WMAP data release, reporting an upper tail probability even lower than for the 1-year data.

As an illustration of the HC statistic, in figure~\ref{fig:hcis} I represent (in solid red) the $HC^i_{\npix}$ values obtained from the QVW map in the real space, and (in dot-dash blue) the corresponding curve for $R=300$ arcmin. These quantities are normalized to their minimum values, for a better comparison. They are represented in an increasing order. It is remarkable that, for the case of the analyses performed on the SMHW coefficients, there is a tail of very large values of $HC^i_{\npix}$, that are not present for the real space case. This is due to the ability of the SMHW transform to enhance features of a given scale and shape. In figure~\ref{fig:hci}, these values are represented on the celestial sphere (left panel for the real space case, and right panel for the SMHW coefficients). The figure indicates that there are not particular signatures in the real space, whereas the SMHW coefficients at $R=300$ arcmin allow us for a clear identification of the features causing the anomalous values of the $HC_\npix (R)$. In particular, the key role played by the \cs is highlighted.

\section{The characteristics of the \csp}
\label{sec:chare}
In this Section, I summarize briefly some of the most important properties of the \csp. I will focus in two major aspects: its morphology and its frequency dependence.

The morphological properties of the \cs are different depending whether we do the analysis in the real or wavelet space. As it was pointed out in~\cite{cruz05}, the region associated with the \csp, in the real space, appears as formed by several small cold spots. The amplitude of the most prominent of these spots is $\lesssim - 350\mu K$ with a size of $\approx 1^{\circ}$. None of these structures is particularly anomalous. An image of the \cs in the real space is shown in the left panel of figure~\ref{fig:patch}. It is, however, in the wavelet space where the \cs appears more interesting. In the right panel of figure~\ref{fig:patch}, I present a close view of the \cs after convolution with the SMHW at a scale $R=250$ arcmin. 
Besides all the anomalous characteristics previously discussed (i.e., area, HC, and $\mas$), the \cs appears as a very symmetric feature. However, this effect could be biased since, after all, the SMHW is an isotropic filter and, therefore, the symmetric features of the \cs could be amplified, erasing any possible intrinsic anisotropy.
This issue was studied in detail by~\cite{cruz06}. Instead of applying an isotropic wavelet, the anisotropic Elliptical Spherical Mexican Hat Wavelet (ESMHW) was adopted. The cleaned CMB map derived form the WMAP was transformed into ESMHW coefficients (at the scales for which the WMAP data appeared as anomalous), for different ratios $\zeta$ between the smallest and the largest axes of the ESMHW, and for different orientations. This work proved that the maximum matching between the \cs and the ESMHW took place when $\zeta \in \left[0.875, 1\right]$ and, hence, indicated that the \cs structure was quite close to be isotropic (assuming that the ratio of the ESMHW axes mimics, somehow, the ratio between the \cs \emph{axes}).
%%%
\begin{figure*}
\begin{center}
\includegraphics[width=8cm,keepaspectratio]{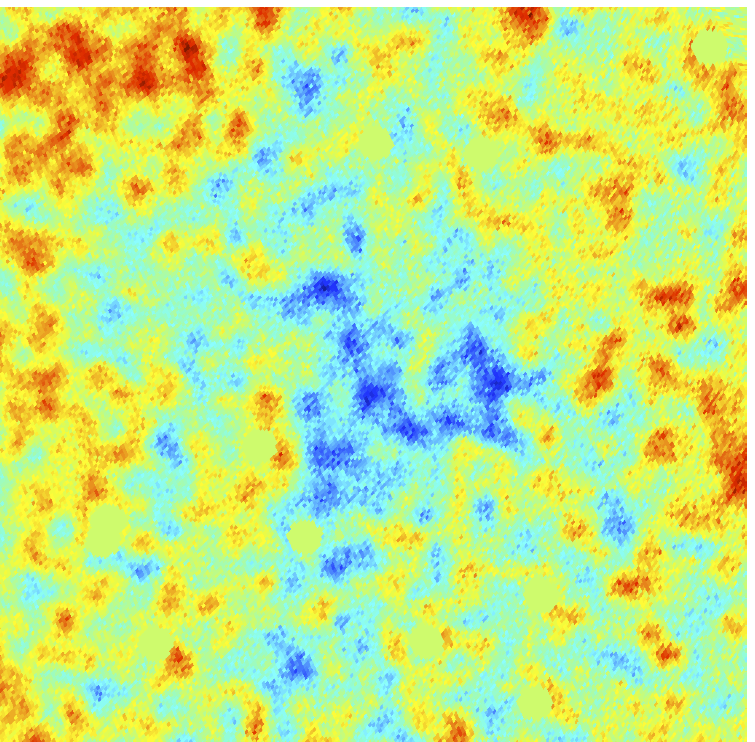}
\includegraphics[width=8cm,keepaspectratio]{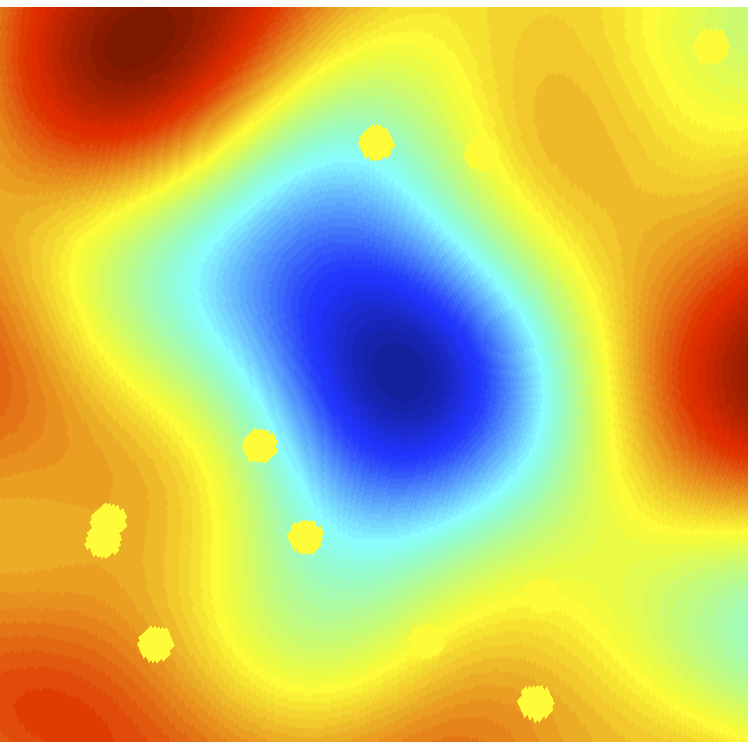}
\caption{~\label{fig:patch}Image of the \cs ($\theta=-147^\circ,\phi=209^\circ$) in the real (left panel) and wavelet (right panel) spaces. The homogenously filled circles correspond to positions where known extragalactic point sources have been masked.}
\end{center}
\end{figure*}
%%%%%

The frequency dependence of the \cs has been a matter of study soon after its discovery. Already in~\cite{vielva04} the SMHW coefficients of the cleaned WMAP frequency channels (namely, Q, V and W bands) were computed, and the mean value of the coefficients associated with the \cs at the scale $R=250$ arcmin was estimated.
No obvious frequency dependence of this mean value of the wavelet coefficients was found, hence, being fully consistent with the expected behaviour for the CMB emission\footnote{The analyzed WMAP frequency channels were in thermodynamical temperature and, therefore, the CMB appears as a frequency independent emitter. Notice that, since the SMHW transform is a linear operation, the same behaviour is expected for the wavelet coefficients.}, and, therefore, quite different to the typical frequency dependence of the Galactic foregrounds.

\section{The significance of the detection: the \emph{a posteriori} issue}
\label{sec:sig}
One of the most questioned aspects of the \emph{WMAP anomalies} in general, and of the \cs in particular, is the issue of the actual significance of the detection. This is a very important point that is intimately linked to the \emph{blind} nature of all the Gaussianity/isotropy tests that led to the report of such anomalies. 

Let me review where this problem comes from: if many tests are performed in a given data set, it is not strange that some of them report some deviation from the null hypothesis. 
It is quite usual to face the following situation: a set of \emph{blind} tests (i.e., tests that just challenge the compatibility of the data with a given null hypothesis, $\mathrm{H_0}$, and not confronting such hypothesis with an alternative one, $\mathrm{H_1}$) claim a given incompatibility of the WMAP data.
A subsequent test is performed, taking into account the previous finding and, usually, in such a way that the initial reported deviation is now found at higher significance. In this procedure there are two weak points: the first one, already mentioned, is to assess the probability of finding a deviation as the one claimed during the first step, taking into account all the possible tests that were performed. The second one is the \emph{credibility} of the probability for the follow-up test, where a particularity was studied in greater detail. 

As mentioned above, this is a common situation for the \emph{WMAP anomalies} works and, therefore, the \cs is not an exception. Several tests were made in the first work by~\cite{vielva04}, namely, the estimation of the skewness and the kurtosis at several scales of the SMHW. A particular deviation was highlighted: the excess of kurtosis at several scales around $R=250$ arcmin. After that, the \cs was identified as a prominent feature, and further tests (the $\mas$, the cold area, the HC) were applied.
I believe that there is not a unique and clear way to solve these ambiguities and, to my view, this point is not usually addressed in the literature. However, whereas for the latter aspect (i.e., the significance for the follow-up tests) the solution is hard, for the former there could be some possible getaway, at least depending on the complexity of the preliminary analysis. Actually, the \cs is one of the few \emph{WMAP anomalies} where this particular aspect has been considered with deeper interest. In fact, it was the matter of several papers~\cite{mcewen05,cruz06} and, in particular, of~\cite{cruz07a}. 

In this last work, the significance of the first detection was addressed, focusing in the \emph{a posteriori} selection of the statistic (the kurtosis) and the scale range ($\approx 250$ arcmin). A conservative procedure to establish the p-value of the non-Gaussian detection, based on the characteristics of the analysis, was proposed. 
More specifically, since 30 statistics were applied to the QVW map (i.e., the skewness and the kurtosis of the SMHW coefficients at 15 scales), and only 3 out of these 30 statistics were found as anomalous (i.e., the kurtosis of the SMHW at scales $R=200, 250$, and, $300$ arcmin were outside the 1\% acceptance interval |see ~\cite{cruz07a} for details), then, it was decided to estimate the significance of the non-Gaussianity detection by exploring in how many out of 10,000 CMB Gaussian simulations it was observed that the skewness or the kurtosis of the SMHW coefficients were outside the 1\% acceptance interval, at least, at three scales. The p-value obtained in this manner was 0.0185. This p-value can serve, as explained in~\cite{cruz07a}, as a conservative probability related to the non-Gaussianity associated with the SMHW analysis.

In this spirit, the following up tests (e.g., the amplitude, the area or the HC) can be just seen as additional probes to explore/understand the previous deviation, rather than as independent sources for establishing a proper significance level for the detection.

Recently,~\cite{zhang10} has questioned the non-Gaussianity found by~\cite{vielva04}, since the excess of the kurtosis was clearly found with the SMHW, but it was not the case with other analyzing kernels (proposed in~\cite{zhang10}), as the top-hat and the Gaussian filters. 
The authors argued that these tools are more \emph{natural} than a wavelet like the SMHW and that, therefore, the selection of the SMHW is somehow
\emph{a posteriori}. Contrary to this reasoning, I found that the results obtained by~\cite{zhang10} imply a different conclusion: the lack of detection when analyzing with the top-hat and the Gaussian filter is a proof of the issue discussed in Section~\ref{subsec:wave}, namely, that any filtering kernel is not necessary suitable for the detection of any non-Gaussian feature. It is clear that some features (like point sources, cosmic strings or textures) are much better detected after applying optimal or targeted filters, rather than general ones (like the top-hat or the Gaussian functions). 

The reason why a compensate filter as the SMHW gets a much larger amplification as compared to uncompensated kernels as the previous ones is that it is much more efficient to remove the background fluctuations above and below a given scale interval. Even more, it can be shown (e.g.,~\cite{herranz02})  that the SMHW is close to the optimal or \emph{Matched} filter to detect objects with a Gaussian-like profile embedded in CMB-dominated background (as it is the case in the region of the \csp). The form of a \emph{Matched filter} designed for a given situation is defined, not only by the shape of the feature to be detected, but also by the statistical properties of the background. In particular, the \emph{Matched filter} (defined in the Harmonic space) is proportional to the shape of the feature, and inversely proportional to the angular power spectrum of the background. Therefore, bearing in mind that (at degree scales) the CMB is well described by an angular power spectrum close to $C_\ell \propto \ell^{-2}$, it is trivial to show that the SMHW is near to be an optimal tool for detecting a feature described by a Gaussian-like profile, and embedded in such background.

Let me remark that, as it was said in Section~\ref{subsec:wave}, the selection of wavelets as a suitable tool for non-Gaussian analysis, as it was shown by several authors in the past (e.g., \cite{hobson99,aghanim99,martinezgonzalez02}), can not be considered as an \emph{a posteriori} choice, but, rather, as a natural option for studying scale-dependent phomena. In particular, the compensation property satisfied by wavelets make them extremely good analyzing kernels to amplify certain features, precisely because it assures a strong suppression of the large scale fluctuations of the background. 

%%%
\begin{figurehere}
\begin{center}
\includegraphics[width=8cm,keepaspectratio]{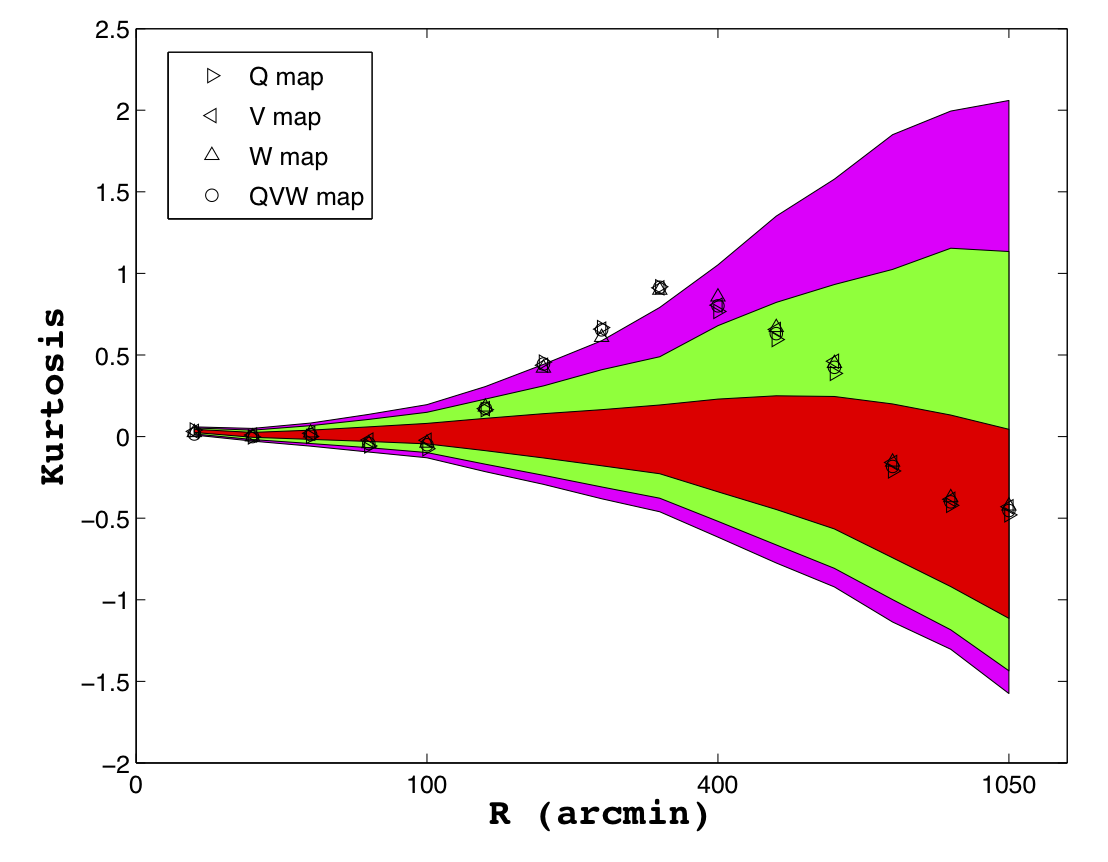}
\caption{~\label{fig:fore_kurto}Frequency dependence of the excess of the kurtosis of the SMHW coefficients. The plot shows the variation of the kurtosis as a function of the scale $K_R$ for cleaned CMB maps obtained from the WMAP data at three cosmological frequencies (Q-,V-, and W-bands), and
for the optimal CMB cleaned map provided by WMAP, as noise-weighted combinations of the previous maps (the QVW map). The coloured regions represent, as for other figures, the 32\%, 5\% and 1\% acceptance intervals provided by simulations.}
\end{center}
\end{figurehere}
%%%% 

Finally, it is worth recalling that the \cs has been identified as an anomalous feature by other tools different from the SMHW: by directional wavelets~\cite{mcewen05}, scalar indices~\cite{rath07,rossmanith09}, steerable wavelets~\cite{vielva07a}, needlets~\cite{pietrobon08}, and the Kolmogorov stochastic parameter~\cite{gurzadyan09}.

\section{Some possible sources to explain the \csp}
\label{sec:sources}
To find an explanation for the non-Gaussianity deviation associated with the \cs is the next step, once its anomalous nature (i.e., non-compatible with the standard inflationary scenario) is accepted.
With this aim, many efforts have been done in the last years, considering different sources for the observed anomaly. The possible causes addressed so far account for \emph{systematics} effects, mostly due to instrumental aspects that are not well understood/modelled; spurious emissions due to \emph{foregrounds} or contaminants of the cosmological signal; non-accounted \emph{secondary anisotropies} induced on the CMB photons, as the interaction with the ionized medium (e.g., the Sunyaev-Zeldovich effect) or the non-linear evolution of the gravitational potential (e.g., the Rees-Sciama effect); and, of course alternative (or complementary) models to the standard inflationary scenario (as cosmic defects).
In the following subsection, and also in Section~\ref{sec:texture}, I address these possibilities, starting from those hypotheses that are less dramatic from the point of view of strong implications for the standard cosmological model.

%%%
\begin{figure*}
\begin{center}
\includegraphics[width=5.8cm,keepaspectratio]{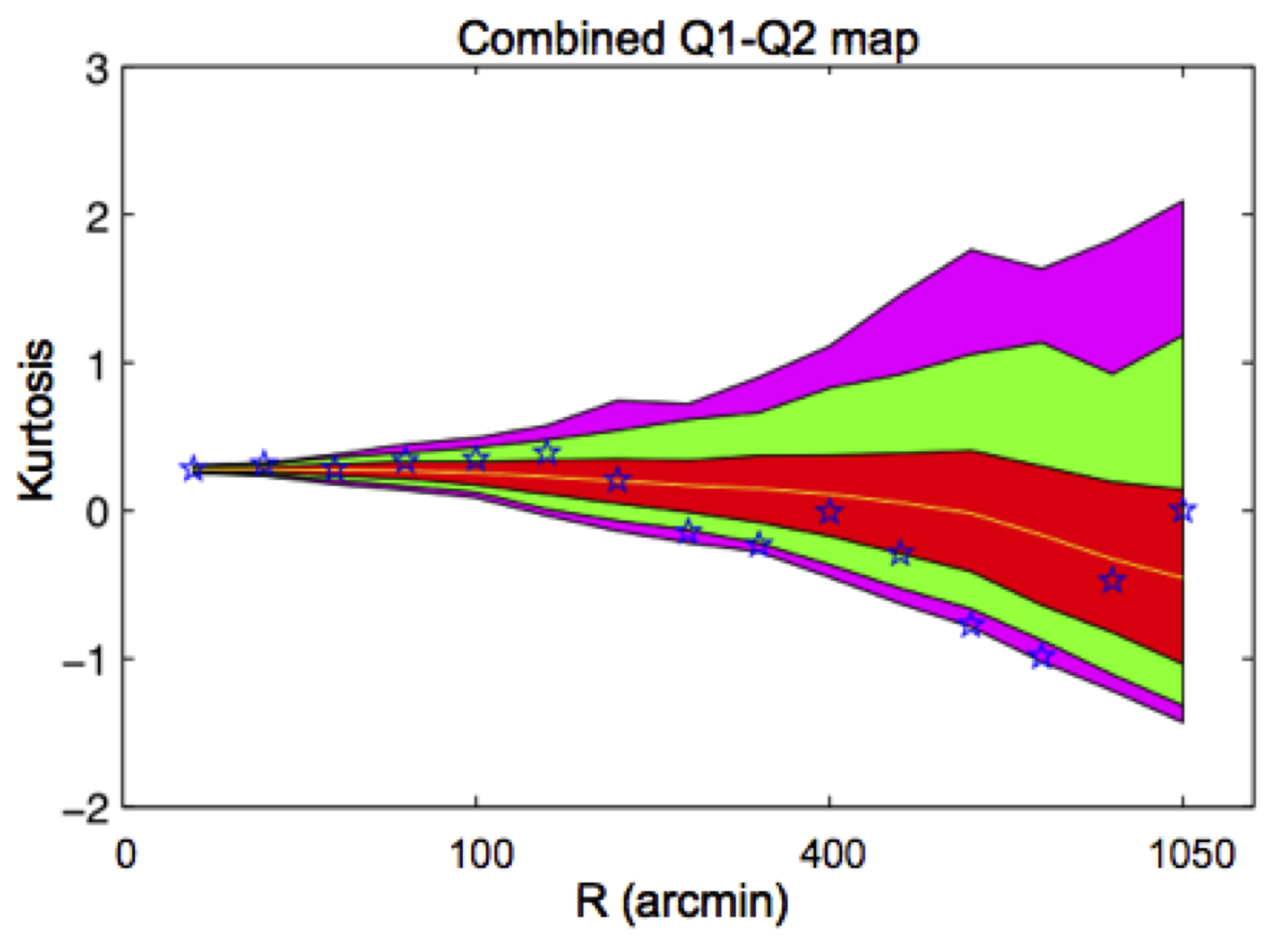}
\includegraphics[width=5.8cm,keepaspectratio]{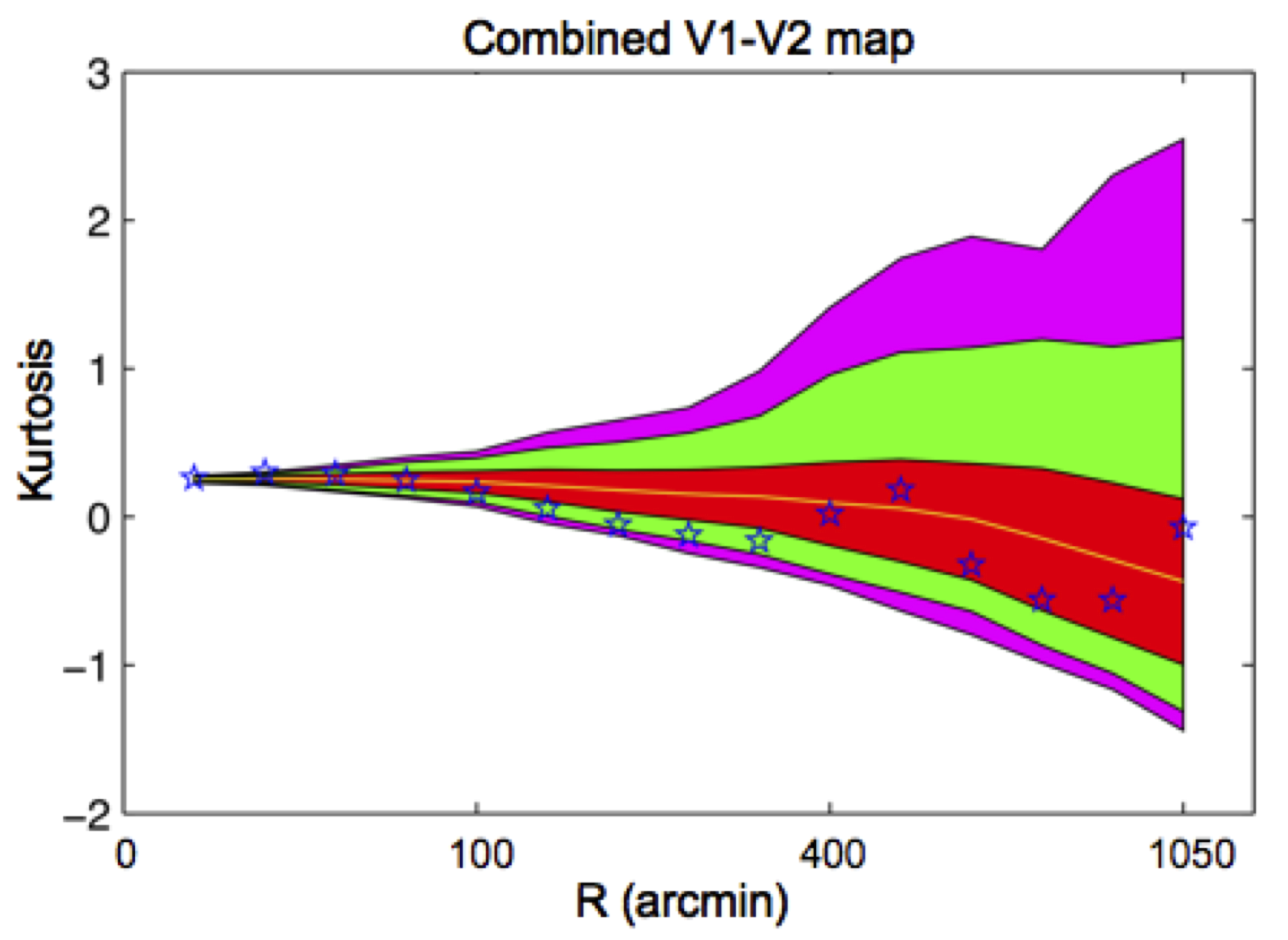}
\includegraphics[width=5.8cm,keepaspectratio]{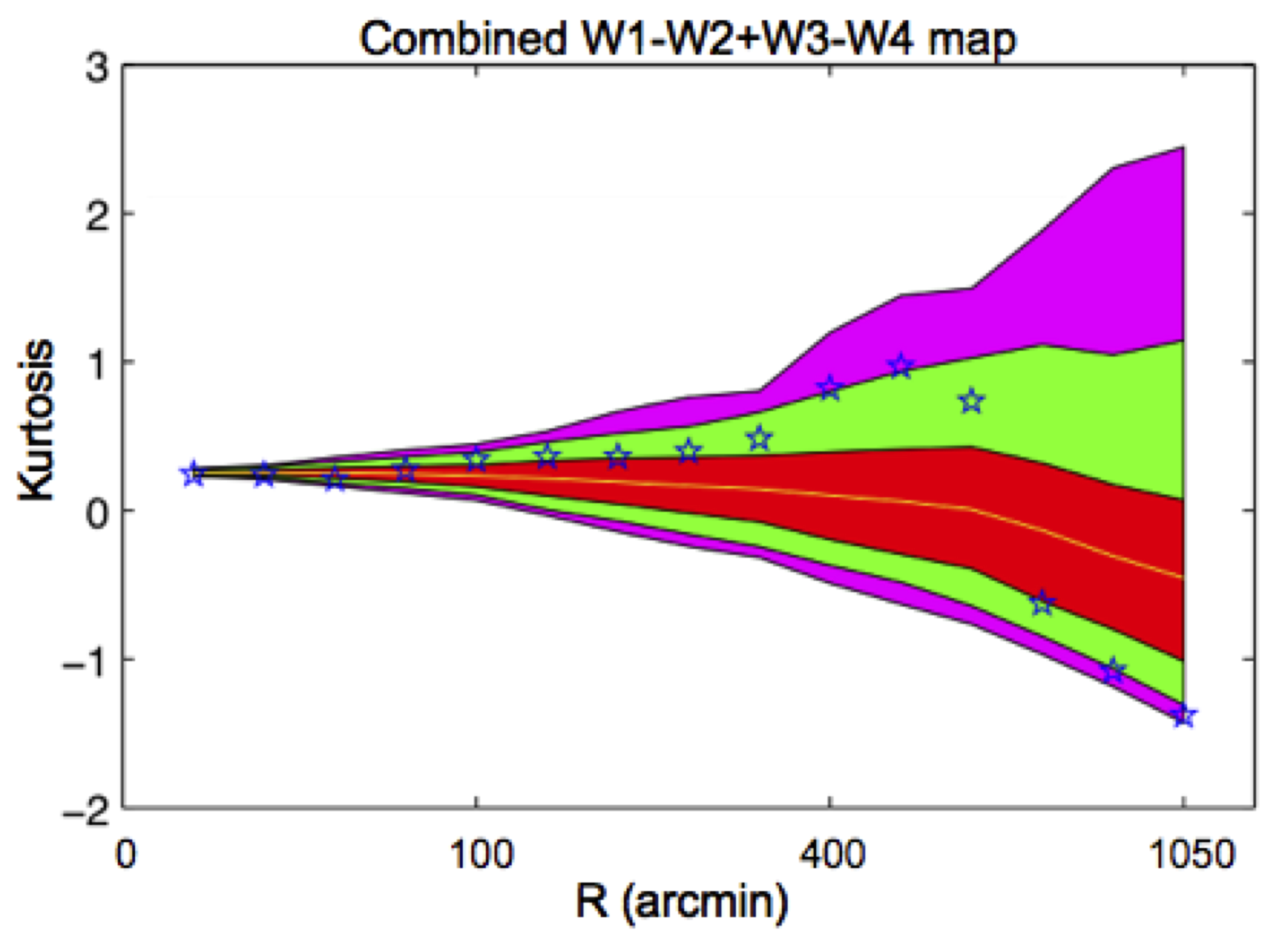}
\caption{~\label{fig:sys}\emph{Null} tests performed on difference maps, free of any foreground and cosmological signal. The left panel represents the kurtosis of the SMHW coefficients, as a function of the scale $R$, for the $Q1-Q2$ map, i.e., for a map built as the difference of the difference assemblies observations at 41 GHz. The middle panel is as the previous one, but for the $V1-V2$ difference map, i.e., at 61 GHz. Finally, the right panel provides the variation of the SMHW coefficients as a function of the wavelet scale, but for the difference map obtained with the 4 difference assemblies of WMAP at 90 GHz, i.e., $W1-W2+W3-W4$. The blue asterisks represent values for the data, whereas red, green and magenta regions represented the acceptance intervals at 32\%, 5\% and 1\%, respectively, determined by simulations.}
\end{center}
\end{figure*}
%%%%

\subsection{Systematics}
\label{subsec:sys}

To test the influence of unknown/unmodeled systematics on the non-Gaussianity deviation is, as one could imagine, a very hard task. The analyses that can be carried out to probe such sources are, basically of two types. 

One of these analyses are \emph{consistency tests}. As most of the CMB experiments, WMAP satellite can provided us with cleaned CMB maps (for instance, following the template fitting approach described by~\cite{gold09}) for several detectors. Therefore, focusing in the non-Gaussian deviation associated with the \csp, an obvious procedure would be to check whether the application of the different statistical tools reveals that such a feature is associated with only one detector, or a smaller set of detectors. 
If this were the case that would be a clear indication for a lack of consistency and, therefore, that the non-Gaussianity detection is associated with a given instrumental feature. This was done by~\cite{vielva04} and~\cite{cruz05} for the kurtosis ($K_R$) and the area ($A^{\nu}_R$) of the SMHW wavelet coefficients. No inconsistency was found: the excess of kurtosis and of area was found to be the same for every difference assembly. As an illustration, in~\ref{fig:fore_kurto} the kurtosis of the SMHW coefficients (as a function of the scale $K_R$) for 4 different CMB maps is presented. In particular, results for the Q-, V-, and W-band cleaned CMB maps are shown, together with the curve obtained from the analysis of the optimal QVW-map. These curves are quite similar, which indicates that the non-Gaussian signal is presented in all the WMAP detectors, at a similar level.

The second type of analyses are \emph{null} tests. The cosmological frequencies of the WMAP satellite (i.e., Q band at 41 GHz, V-band at 61 GHz and W-band at 94 GHz) are made from more than one difference assembly. Hence, just subtracting difference assemblies at the corresponding band can produce noise maps per frequency. Neglecting small differences from the optical beams and the band-pass widths, the CMB and foreground emissions have been cancelled out in this new map. 
Therefore, the application of the statistical tools to these difference maps helps to check whether the non-Gaussian signal is a \emph{noisy artifact} (if such signal is still present) or not (if consistency with Monte Carlo simulations is found).
These was done by~\cite{vielva04} and~\cite{cruz05}, again, for the kurtosis and the area of the SMHW coefficients, respectively. As for the previous type of systematics probe, there was a clear indication that the non-Gaussian signal was not related to any instrumental signature. As an example, in figure~\ref{fig:sys}, the result obtained by~\cite{vielva04} for $K_R$ is presented. On the left panel, I represent the variation of the SMHW wavelet coefficients for the difference map constructed at 41 GHz as $Q1-Q2$. Similarly, the results for the $V1-V2$ map at 61 GHz are provided in the middle panel. Finally, in the right panel, I give the output for the difference map obtained at the 94 GHz band as the combination $W1-W2+W3-W4$. Results for the WMAP data is given as blue asterisks, whereas, as for previous figures, the red, green, and magenta regions provide the 32\%, 5\% and 1\% acceptance intervals, respectively.

Summarizing, \emph{consitency} and \emph{null tests} do not reveal the presence of systematics behind the non-Gaussianity associated with the \cs. Besides these test, it is important to remark that the angular size associated with this feature is $\approx 10^\circ$ in the sky. It is not trivial to think in a systematic effect affecting at this scale, and providing such localized feature in the sky as the \cs.
 
Finally, the subsequent releases of the WMAP data (where the modelling of the instrumental properties have been improving with time) have shown that there are not changes in the non-Gaussianity deviation, except for a slight increasing on its significance, which reflects the higher signal-to-noise that WMAP data is getting as observational time increases.

\subsection{Foregrounds}
\label{subsec:fore}
Astrophysical contaminants or \emph{foregrounds} are the next possible origin for the non-Gaussianity associated with the \csp,. It is well known that foregrounds are highly non-Gaussian signals. It is worth commenting that, although the \cs is negative, it is still possible to think in an \emph{additive} source (as foregrounds are) as a feasible explanation. To understand this point, it is important to recall that the 
QVW map (that, as I said before, is commonly adopted in the literature for cosmological analyses) is obtained as a noise-weighted linear combination of cleaned CMB maps at different frequencies. These maps (at 41, 61 and 94 GHz) are produced, as mentioned previously, via a template fitting (see, for instance,~\cite{gold09}). Therefore, any over-subtraction of foregrounds templates could caused a \emph{foreground residual} in the form of a \emph{cold} emission\footnote{In some works, a VW map is adopted, i.e., a map built as a noise-weighted linear combination of the cleaned V- and W-bands.}.

Since, as mentioned before, the size of the \cs is of several degrees, it is really hard to believe that this feature could be associated with residuals from point sources (i.e., from radio and infrared galaxies). Therefore, only galactic emissions (as synchrotron, free-free, and thermal and spinning dust) could be responsible for a large feature as the \csp. However, notice that the \cs is located at $57^\circ$ from the Galactic plane, and at a longitude of $\ 209^\circ$. In other words, the \cs is placed in a region of low Galactic contamination.
According to the previous reasoning, it is already hard to make compatible the presence of the \cs with a given Galactic emission. Even though, of course, the issue has been a matter of discussion. I review here some of these analyses.

The most obvious test is, of course, to check whether there is or not any frequency dependence of the statistical estimators that indicated the Gaussian deviation. For instance, in~\cite{vielva04,cruz06}, the kurtosis of the SMHW coefficients ($K_R$) was studied for the different CMB cleaned maps obtained at the Q-, V-, and W-bands. 
%%%
\begin{figurehere}
\begin{center}
\includegraphics[width=8cm,keepaspectratio]{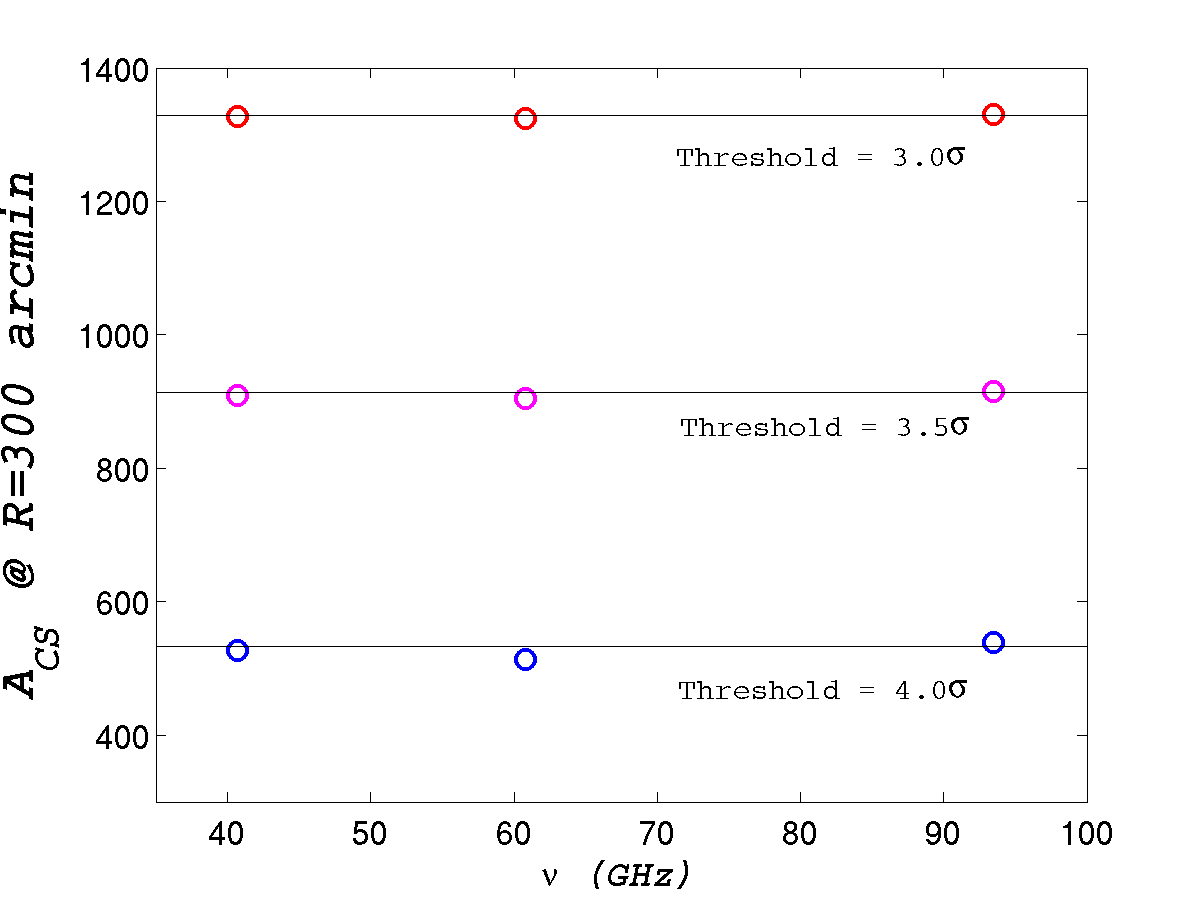}
\caption{~\label{fig:fore_area}Frequency dependence of area of the SMHW coefficients. The plot shows the area of the wavelet coefficients at $R=300$ arcmin for different thresholds (from top to bottom: 3, 3.5, and 4$\sigma_R$), as a function of the frequency. The circles represent the values obtained for cleaned CMB maps obtained from the WMAP data at three cosmological frequencies (Q-,V-, and W-bands). The solid lines represent the areas, at different thresholds, for the noise-weighted QVW map.}
\end{center}
\end{figurehere}
%%%%

The results obtained for this study are given in figure~\ref{fig:fore_kurto}.
The kurtosis $K_R$ is presented for the 41, 61, and 94 GHz channels, and for the noise-weighted lineal combination (the QVW map). It is remarkable the high similarity of the curves. The pattern of the kurtosis, as a function of the SMHW scale, is the same for all the maps. The same is observed for its normalization. An equivalent test can be done for the area of the SMHW coefficients ($A^{\nu}_R$), as proposed by~\cite{cruz05,cruz06}. Results, at $R=300$ arcmin, are presented in figure~\ref{fig:fore_area}. Notice that the agreement of the area of the SMHW coefficients (above threshold $\nu =$3, 3.5, 4$\sigma_R$) among different frequency bands (Q, V, and W) and the combined QVW map (represented by the solid lines in the figure) is very high. These kind of tests show that there is not any evident frequency dependence of the statistics associated with the non-Gaussian deviation and, therefore, that such anomaly is fully consistent with the expected behaviour for a CMB feature.  

Additional tests supporting this idea have been proposed in the literature. First~\cite{vielva04}, the kurtosis of the SMHW coefficients was analyzed for a CMB-free map, constructed as the combination of the 4 difference assemblies at the W-band minus the sum of the 4 ones at Q- and V-band (i.e., $W1+W2+W3+W4-Q1-Q2-V1-V2$). This kind of map could have a contribution of the Galactic contaminants (outside a given observing mask), since foreground emissions are not frequency independent. This analysis did not show any significant deviation from the expected behaviour from Gaussian simulations, and, therefore, discarding the presence of significant foreground residuals. 

Second, as suggested by~\cite{cruz06}, different CMB recoveries from the WMAP data (where independent and alternative component separation approaches were followed) could be analyzed. In particular, the cleaned CMB maps obtained by~\cite{tegmark03} were probed.  The kurtosis, the area, and the $\mas$ of the SMHW coefficients do not changed significantly from the different CMB maps. 

\cite{cruz06} explored a more complicate scenario: a situation in which combinations of different foreground emissions could mimic, in the region of the \csp, the behaviour associated with the CMB, i.e., a frequency independent global emission. To check this possibility, several templates were used as tracers of the Galactic foregrounds, namely, the Rodhes/HartRa0 2326 MHz~\cite{jonas98} radio survey for synchrotron, the H$_\alpha$ by~\cite{finkbeiner03} for the free-free, and the thermal dust model by~\cite{finkbeiner99}. Authors studied the expected contribution of foregrounds in the region of the \csp, and they found that, taking into account the uncertainties in the extrapolation of these templates from their original observations to the WMAP frequency range, it was possible to find a global Galactic emission that was nearly frequency-independent from 41 to 94 GHz (i.e., from Q- to W-bands). However, it was found that the emission was at a level of one order of magnitude below the \cs temperature. It was checked that, even accounting twice for that hypothetical foreground emission, it was not possible to reconcile observations with the Gaussian model.

All these tests on the impact of the foregrounds indicated that the non-Gaussian signal associated with the \cs was fully consistent with a CMB like frequency dependence, and that the role played by astrophysical contaminants was negligible.

\subsection{The Sunyaev-Zeldovich effect}
\label{subsec:sz}
After checking that the possible impact of systematics and foregrounds on the non-Gaussian detection is very low, the next step is to study whether secondary anisotropies of the CMB could explain the anomalous nature of the \csp. 

The Sunyaev-Zeldovich effect\footnote{It is produced by the inverse Compton interaction of the CMB photons,  as they cross the hot electron gas that is found in clusters of galaxies.} (SZ) could be a potential candidate to explain the anomaly. Two major reasons support this possibility: first, the size of the \cs is nearly compatible with the fluctuations caused by the nearest clusters of galaxies, and, second, this fluctuations (in the frequency range covered by WMAP) produce cold spots~\cite{sunyaev70}.
%%%
\begin{figurehere}
\begin{center}
\includegraphics[width=8cm,keepaspectratio]{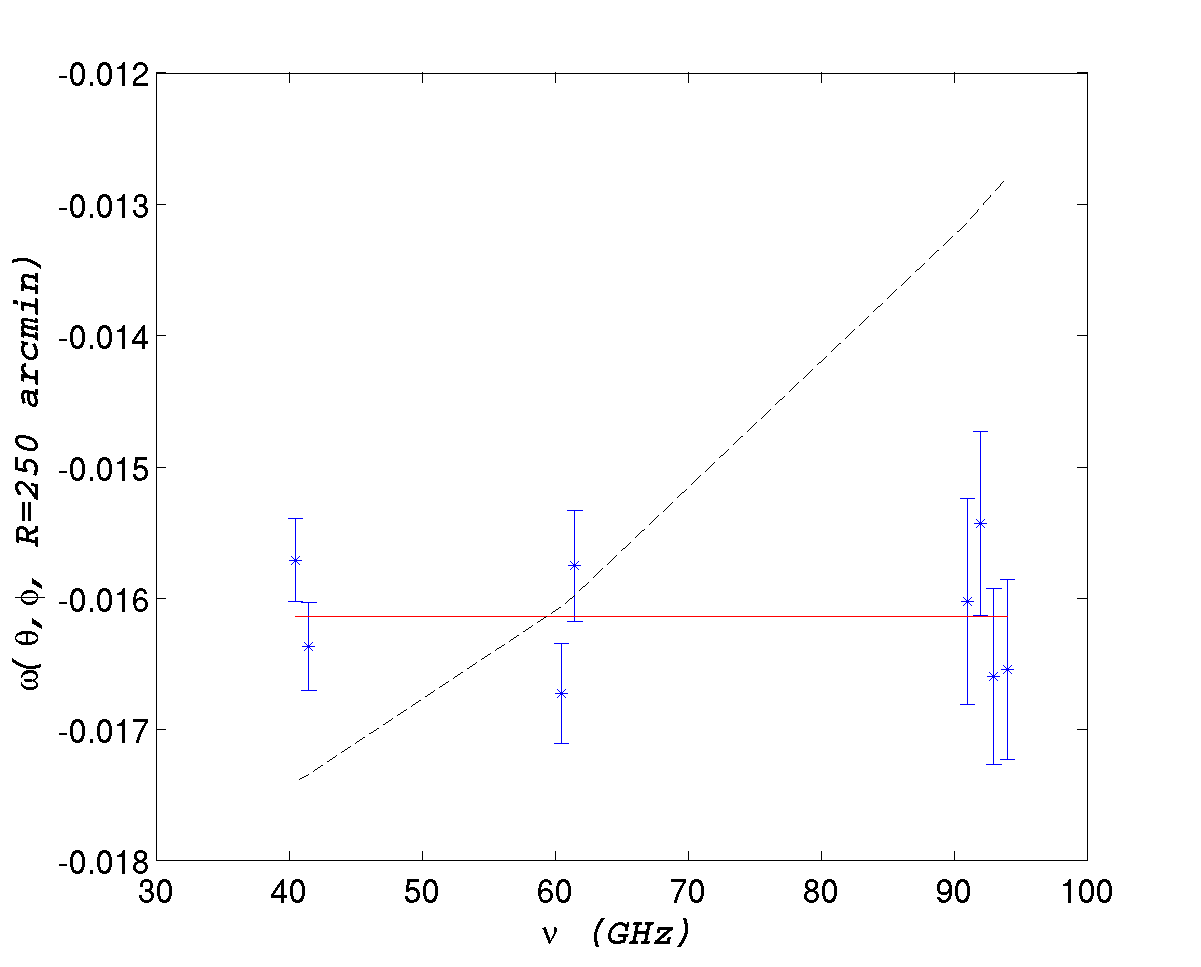}
\caption{~\label{fig:sz}Best-fit of the WMAP data |across the 8 difference assemblies for the Q-, V-, and W- bands| to a CMB (solid line) and a SZ spectra (dashed line). The fit is performed to the minimum amplitude of the \cs in wavelet space, at a scale of $R=300$ arcmin. Error bars are obtained from Gaussian simulations.}
\end{center}
\end{figurehere}
%%%%

There is not evidence for the existence of any large cluster in the direction of the \csp. 
However,~\cite{brough06} reported the presence of a large concentration of galaxies in that direction |referred to as the \emph{Eridanus super-group}|, that could account for a total mass of $\approx 10^{14}M_\odot$ (see~\cite{cruz08} for details).

This scenario implies, therefore, that the SZ could explain the nature of the \csp, at least, partially. In other words, it could be possible that a combination of a SZ contribution plus a large (but not anomalous) CMB fluctuation could account for the \cs emission.
This was studied by~\cite{cruz06}, and the results are summarized in figure~\ref{fig:sz}. In this figure, the values of the SMHW coefficients $\wcr$, at the position of the \cs and at $R=300$ arcmin, are given for the eight CMB cleaned maps at the Q-, V-, and W-bands. The corresponding error bars were computed from CMB plus noise simulations, corresponding to the instrumental properties of these detectors, and convolved with the appropriate SMHW kernel.

Three different fits to the data were explored. First, a pure CMB spectrum was using, giving a very good fit with a reduced $\chi^2 = 1.00$. Such fit is represented by the solid line in figure~\ref{fig:sz}. Second, a pure SZ spectrum was tested, obtaining a very poor fit (dashed line) with a reduced $\chi^2 = 9.12$. Finally, a joint fit to a CMB plus SZ spectra was explored, obtaining an amplitude for the SZ spectrum consistent with zero, and a reduced $\chi^2$ quite similar to the first case. These results confirm that the frequency dependence of the \cs is consistent with a CMB-like spectrum. They also rule out the possibility that the SZ is playing any significant role.

\subsection{The late evolution of the large scale structure}
\label{subsec:rs}
Another secondary anisotropy that could explain the anomalous nature of the \csp, is the one due to the non-linear evolution of the gravitational field: the so-called Rees-Sciama effect (RS). In particular, it is known~\cite{martinezgonzalez90a,martinezgonzalez90b} that voids in the large scale structure could induce a negative non-linear anisotropy in the CMB photons. The size of such secondary anisotropies depends on the proper size of the void and its redshift. Therefore, as for the SZ, the RS is another potential candidate to explain the anomalous nature of the \csp.

Extra support for this hypothesis came from two different paths. 
On the one hand, theoretical works as~\cite{inoue06,inoue07} proposed that a very large void (of $\approx 300h^{-1}$ Mpc) and located at low redshift ($z \ll 1$) could produce large negative CMB fluctuations such as the \csp, even with modest density contrast values (i.e., in a quasi-linear regime). 
On the other hand, it was suggested~\cite{rudnick07} that the NVSS catalogue~\cite{condon98} seems to show, at the position of the \csp, a lack in the number count of radio galaxies. 

Against these ideas, some criticisms can be made. First, this kind of voids are not observed and, even more, according to current N-body simulations~\cite{colberg05} they are extremely rare events ($\approx 13\sigma$, i.e., much more rare than the \cs deviation itself, that was a $98.15\%$ event!) in the standard cosmological framework. Second, the claim made on the NVSS data has been recently questioned by~\cite{smith10}, suggesting that such finding was an artefact caused by possible systematics related with the NVSS data processing, the statistical procedure and the 
\emph{a posteriori} selection of the \cs position. 

In addition, observational campaigns on the region of the \cs were made by~\cite{granett09} with the MegaCam on the Canada-France-Hawaii Telescope, and by~\cite{bremer10} with the VIMOS spectrograph on the Very Large Telescope. 
Both works have reported that the large scale structure in that direction, and up to $z \approx 1$, is fully consistent with the standard model, and that there is no evidence of such large voids as those required by the non-linear evolution of the gravitational potential hypothesis.

\section{A plausible explanation: cosmic textures}
\label{sec:texture}
In the previous Section, I have presented an overview of the several works carried out to establish whether the non-Gaussianity detection associated with the \cs could be explained in terms of systematics, foregrounds, and secondary anisotropies as the Sunyaev-Zeldovich and the Rees-Sciama effects.
None of these possibilities seem to provide a satisfactory explanation and, therefore, other sources should be investigated. 

In this Section, I pay attention to the suggestion made by~\cite{cruz07b}: the \cs could be caused by a \emph{cosmic texture}. Cosmic textures~\cite{turok89} are a type of cosmic defects. They are supposed to be generated at some stage of the early Universe, associated with the symmetry-braking phase transitions that are predicted by certain theoretical models of high energy physics (see, for instance,~\cite{vilenkin00} and references therein for a much more detailed explanation). In short, defects can be understood as space regions of a given phase state, surrounded by a space already in a new phase. In some cases, as for textures, these regions could collapse and, therefore, left an imprint on the CMB photons.

Among the different types of cosmic defects, textures are the most plausible candidate to explain the anomalous nature of the \csp, since the interaction of the CMB photons with the time variation of the gravitational potential, associated with an eventual collapse of the texture, produces spots in the CMB fluctuations~\cite{turok90}. 
Even more, cosmic textures are expected to be a possible source of kurtosis deviation, whereas the expected level of skewness is almost negligible (at least for values of the symmetry-breaking energy scale compatible with current observations). This is caused by the even probability of textures producing cold and hot spots and, therefore, providing a nearly symmetric distribution of temperature fluctuations. In fact, the 
equilateral $f_{NL}$ expected from textures goes as $f_{NL} \approx 1.5\times10^{-100} {\psi^6_0}$~\cite{silvestri09},  where $\psi_0$ is the symmetry-breaking energy scale, measured in Gev. For typical limits in $\psi_0$ imposed by the CMB angular power spectrum analysis (e.g.,~\cite{bevis04,urrestilla08}), the 
expected equilateral $f_{NL}$ is $\approx 10^{-9}$, i.e., a tiny value well below 
the current constraints ~\cite{komatsu10}.

The isotropic shape of the temperature fluctuations related to these spots can be approximated, at least at small angular distances $\vartheta$, as~\cite{pen94}:
\begin{equation}
\label{eq:text_profile}
\frac{\Delta T}{T}\left(\vartheta\right)=\pm\varepsilon\frac{1}{\sqrt{1+4\left(\frac{\vartheta}{\vartheta_c}\right)^2}},
\end{equation}
where $\vartheta$ represents the angular distance from the centre of the spot, and $\vartheta_c$ is a characteristic scale parameter of the spot |that is related to the redshift of the spot and the dynamics of the Universe~\cite{cruz07b}. 
The amplitude $\varepsilon$ is proportional to the symmetry-breaking energy scale $\psi_0$: $\varepsilon = 8\pi^2G\psi^2_0$.
%%
%\begin{equation}
%\label{eq:energy}
%\varepsilon = 8\pi^2G\psi^2_0.
%\end{equation}
%%
It is worth remarking that, according to cosmic texture models, the amplitude $\varepsilon$ is the same for every single spot generated by the collapsing defects.

%%%%
\begin{figurehere}
\begin{center}
\includegraphics[width=8cm,keepaspectratio]{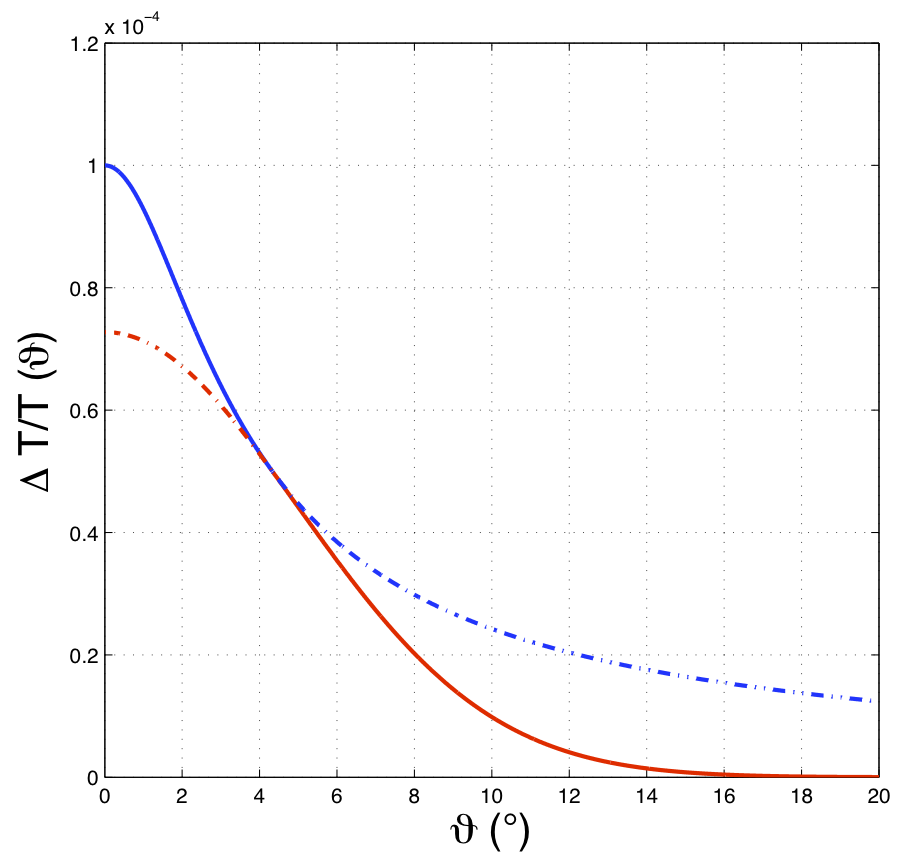}
\caption{~\label{fig:text_profile}Radial profile (solid blue/red lines) of a cosmic texture, as given in equation~\ref{eq:text_profile2}. 
The parameters defining this profile are $\varepsilon = 10^{-4}$ and $\vartheta_c = 5^\circ$. The dashed lines represent the truncated values of the profiles given by equation~\ref{eq:text_profile} (blue), and the Gaussian function (red). See text for details.}
\end{center}
\end{figurehere}
%%%%

As mentioned above, the profile proposed in equation~\ref{eq:text_profile} is only valid up to small relative distances (e.g., $\vartheta \approx \vartheta_c$)\footnote{To my knowledge, there is not any analytical or numerical solution for the full profile of a given cosmic texture yet.}. In order to have a profile valid at larger angular distances,~\cite{cruz07b} proposed to extend the profile given by equation~\ref{eq:text_profile} from its half-maximum, following a Gaussian function. The extension is done by imposing continuity, both, of the profile itself and of its first derivative. 

Taking into account these conditions, the assumed profile for the CMB temperature fluctuation caused by a collapsing cosmic texture would be given by:
\begin{equation}
\label{eq:text_profile2}
\frac{\Delta T}{T} = \pm \left\{
\begin{array}{ll}
\frac{\varepsilon}{\sqrt{1+4\left(\frac{\vartheta}{\vartheta_c}\right)^2}} & \mathrm{if}~~ \vartheta \leq \vartheta_* \\
&\\
&\\
\frac{\varepsilon}{2}\mathrm{e}^{-\frac{1}{2\vartheta^2_c}\left(\vartheta^2 + \vartheta^2_*\right)} & \mathrm{if}~~ \vartheta > \vartheta_*, \\
\end{array}
\right.
\end{equation}
where $\vartheta_* = \sqrt{3}/2\vartheta_c$. 

In figure~\ref{fig:text_profile}, I show the radial section of the above profile (solid line). The red and blue parts of the solid line correspond to the Gaussian function and to the profile of equation~\ref{eq:text_profile}, respectively. The dashed lines represent the truncated regions for both curves. The cosmic texture parameters used in this profile are $\vartheta = 5^\circ$ (i.e., similar to the SMHW scale at which the \cs appears as anomalous) and $\varepsilon = 10^{-4}$. This value corresponds to a symmetry-breaking energy scale of $\psi_0 = 1.13\times 10^{16}$ GeV, that corresponds to a conservative upper limit imposed by CMB measurements (e.g.~\cite{durrer99,bevis04}).

Let me remark that the results obtained with a profile as the one given in equation~\ref{eq:text_profile2}, and that are reviewed in the next Section, do not depend very much with the specific function adopted for the extrapolation. Similar results are obtained, for instance, when an exponential function or a SMHW-like kernel are used.

\subsection{The Bayesian framework}
\label{subsec:bayes}
Attending to the issues discussed in the previous subsection, a cosmic texture could be a strong candidate to explain the \csp: textures produce spots on the CMB temperature fluctuations, they are non-Gaussian signals, and, depending on their amplitude (or the symmetry-breaking scale), they could be compatible with current constraints on the role played by cosmic defects on the structure formation and evolution of the Universe.

\cite{cruz07b} proposed to make use of the texture profile of equation~\ref{eq:text_profile2} to perform a hypothesis test to decide whether the WMAP data (in the position of the \csp) is more likely to be described by a large (but not anomalous) CMB spot (i.e., the \emph{null} $H_0$ hypothesis) or by a cosmic texture of amplitude $\varepsilon$ and size $\vartheta$ added to a random Gaussian and isotropic CMB field (i.e., the \emph{alternative} $H_1$ hypothesis).

The optimal way of performing such hypotheses test is within the Bayesian framework. Bayes' theorem states that, given a data set $D$ and some unknown parameters $\Theta$ (defining a given model in the context of a given hypothesis $H_i$), the posterior probability of the parameters/model given the data
$P(\Theta \mid D,H_i)$, is related to the likelihood $P(D\mid \Theta, H_i)$ (i.e., the probability of the data given the parameters/model) as:
\begin{equation}
\label{eq:bayes}
P(\Theta \mid D,H_i) = \frac{P(D\mid \Theta, H_i)P(\Theta \mid H_i)}{P(D\mid H_i)},
\end{equation}
where $P(\Theta \mid H_i)$ is a measurement of our \emph{a priori} knowledge about the parameters/model (i.e., the prior), and $P(D\mid H_i)$ is a constant (i.e., it does not depend on the parameters/model) called \emph{Bayesian evidence} (BE). The BE is nothing but the average likelihood with respect to the prior,
\begin{equation}
\label{eq:be}
P(D\mid H_i) = \int P(D\mid \Theta, H_i)P(\Theta \mid H_i)\,\mathrm{d}\Theta,
\end{equation}
and it is a largely used mechanism to perform hypotheses test. In particular, its role on different cosmology fields has been quite remarkable during the last years
(e.g.,~\cite{mukherjee06} for dark energy studies,~\cite{bridges07} for anisotropic models of the Universe expansion,~\cite{mukherjee08} for studying different re-ionization models,~\cite{carvalho09} for point source detection, and ~\cite{vielva09} for exploring non-standard inflationary models). 

The importance of BE for hypotheses test is clear. First, it is obvious that the quantity that we would like to obtain is a measurement of the probability of a given hypothesis $H_0$, given the data, i.e., $P(H_0 \mid D)$. This probability can be written, attending to the \emph{probability multiplication rule}, as: 
\begin{equation}
\label{eq:posterior}
P(H_0 \mid D) = P(D \mid H_0)\frac{P(H_0)}{P(D)},
\end{equation}
i.e., it is proportional to the BE and to the probability of the hypothesis, and inversely proportional to the probability of the data.
Under certain circumstances, the probability of the hypothesis could be known, but, however, it is not the case for the probability of the data. In other words, we only can learn about the probability of the hypothesis $H_0$ up to a factor. Therefore, what we can extract is a \emph{relative} measurement of the probability of two hypotheses ($H_0$ and $H_1$), given the same data set $D$. This relative measurement is called the \emph{posterior probability ratio}, $\rho$, and reads:
\begin{equation}
\label{eq:posterior}
\rho \equiv \frac{P(H_1 \mid D)}{P(H_0 \mid D)} = \frac{E_1}{E_0}\frac{P(H_1)}{P(H_0)},
\end{equation}
where, for simplicity, I re-write the BE, $P(D\mid H_i)$, as $E_i$. Hence, if $\rho > 1$, we can conclude that the hypothesis $H_1$ is favoured by the data with respect to $H_0$. In some cases, there is not a clear choice for the probability of the hypotheses. In this case, empirical rules for the ratio of evidences |as the Jeffreys' scale~\cite{jeffreys61}| are usually adopted.

Therefore, the procedure required to explore whether the \cs is more likely to be explained in terms of a cosmic texture ($H_1$) rather than by a Gaussian CMB fluctuation ($H_0$) is clear: first, the likelihood is computed for both hypotheses; second, the BE is estimated, taking into account the adequate priors (equation~\ref{eq:be}); finally, the posterior probability ratio (equation~\ref{eq:posterior}) is evaluated, making use of suitable a priori probabilities for $H_0$ and $H_1$.

This was the procedure followed in~\cite{cruz07b}. It is straightforward to show that, since the \emph{noise term} is caused by standard CMB Gaussian fluctuations and instrumental Gaussian noise, the likelihood  function reads as:
\begin{equation}
\label{eq:likelihood}
P(D\mid \Theta, H_i) \propto \exp{\left(-\chi^2/2\right)},
\end{equation}
where $\chi^2 = \left(D-T(\Theta)\right)C^{-1}\left(D-T(\Theta)\right)^\mathrm{T}$. The correlation matrix $C$ accounts for the full Gaussian
CMB and noise correlations |i.e., $C=S+N$, where $s_{ij} \propto \sum C_\ell P_\ell(\cos \theta_{ij})$ and $n_{ij} = \sigma^2_i \delta_{ij}$, being $P_\ell$ the Legendre polynomials, $\theta_{ij}$ the angular distance between the pixels $i$ and $j$, $\sigma^2_i$ the instrumental noise contribution to pixel $i$, and $\delta_{ij}$ represents the Kronecker delta. 
$D$ represents the data (i.e., the QVW map) and the function $T(\Theta)$ represents the model behind the hypotheses |i.e., equation~\ref{eq:text_profile2} for $H_1$, and $\equiv 0$ for $H_0$.
Finally, $^\mathrm{T}$ denotes standard matrix transpose.

The priors adopted by~\cite{cruz07b} for the parameters $\Theta \equiv \left(\varepsilon\, ,\vartheta_c \right)$ were chosen attending to observational constraints and cosmic texture simulations. In particular, the prior on the amplitude was $\vert \varepsilon \vert \le 10^{-4}$, whereas the prior of the size $1^\circ \le \vartheta_c \le 15^\circ$ was assumed. 
The amplitude prior was uniformly distributed, and, as mentioned above, it is a conservative constraint imposed from the contribution of cosmic defects to the CMB angular power spectrum. 
The size $\vartheta_c $ follows a scale-invariant law, and the limits comes form texture simulations. Textures below $1^\circ$ should be smeared out by photon diffusion and, in addition, they would be related to collapsing events above redshift $\approx 1000$, which would not affect the CMB image. The upper limit is due to the unlikely probability of generating such large textures in the finite celestial sphere. Even so,~\cite{cruz07b} tested that results were not specially sensitivity to the prior selection, since the likelihood was clearly peaked, within a region of the parameter space clearly allowed by observations and texture models.

The marginalization of the posterior probability in equation~\ref{eq:bayes}, led to the determination of the texture parameters, obtaining 
$\varepsilon = 7.3^{+2.5}_{-3.6}\times10^{-5}$ and $\vartheta_c = {4.9^\circ}^{+2.8^\circ}_{-2.4^\circ}$ at 95\% confidence. The 
BE ratio was 150, which, in terms of the empirical rules~\cite{jeffreys61}, is a \emph{strong} indication that the texture hypothesis for the \cs is favoured over the isotropic and Gaussian CMB fluctuation option. Adopting a ratio for the probability of the hypotheses given by the fraction of the sky that is covered by a cosmic texture as large as the one required for the \cs ($\approx 0.017$), the posterior probability ratio was $\rho = 2.5$, which also favours the texture hypothesis.

It is worth mentioning that the estimated value for the texture amplitude could be affected by selection bias. In~\cite{cruz07b} it is established that such bias could provide an overestimate of the texture amplitude by a factor of 2. This bias is caused because the texture amplitude is estimated in a low signal-to-noise regime, were the features placed in large background fluctuations are more easily detected.
Even so, the estimated value for the texture amplitude, $\varepsilon = 7.3\times10^{-5}$, would imply a symmetry-breaking energy scale of $\psi_0 = 8.7\times10^{15}$ GeV, which, on the one hand, is fully compatible with more recent constraints imposed from the analysis of the CMB temperature and polarization power spectra (e.g.~\cite{urrestilla08}), and, on the other hand, is in agreement with the predictions of most of the models for particle physics. In addition, by relating the angular size of the CMB texture profile to the cosmological parameters defining the geometry and evolution of the Universe, it was possible to establish that the texture collapse (that generated the CMB profile in equation~\ref{eq:text_profile2}) occurred at redshift $z\approx6$.

%%%%
\begin{figure*}
\begin{center}
\includegraphics[width=5.8cm,keepaspectratio]{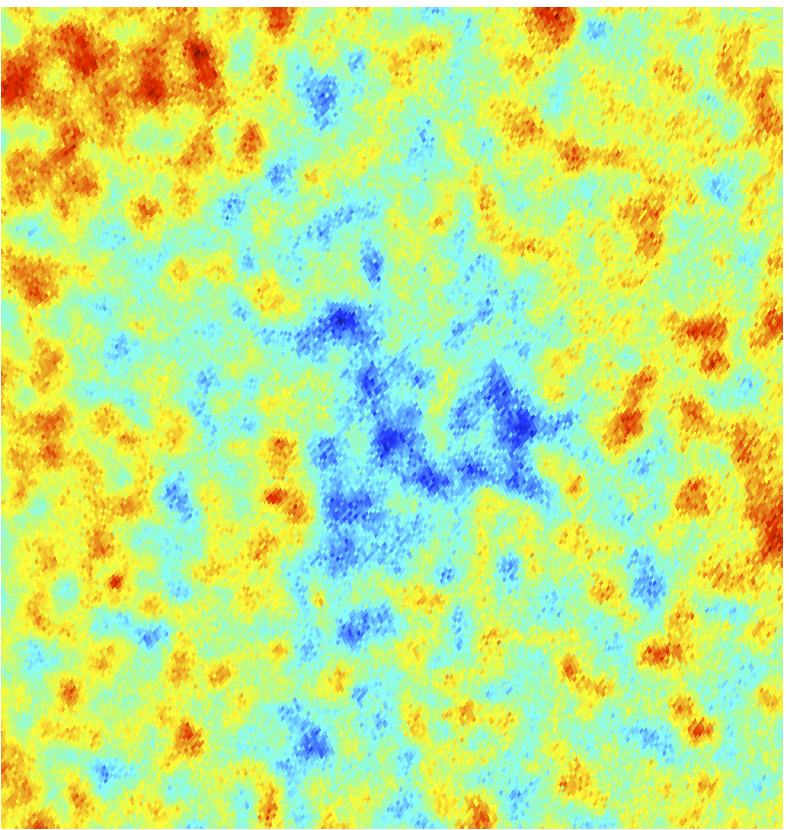}
\includegraphics[width=5.8cm,keepaspectratio]{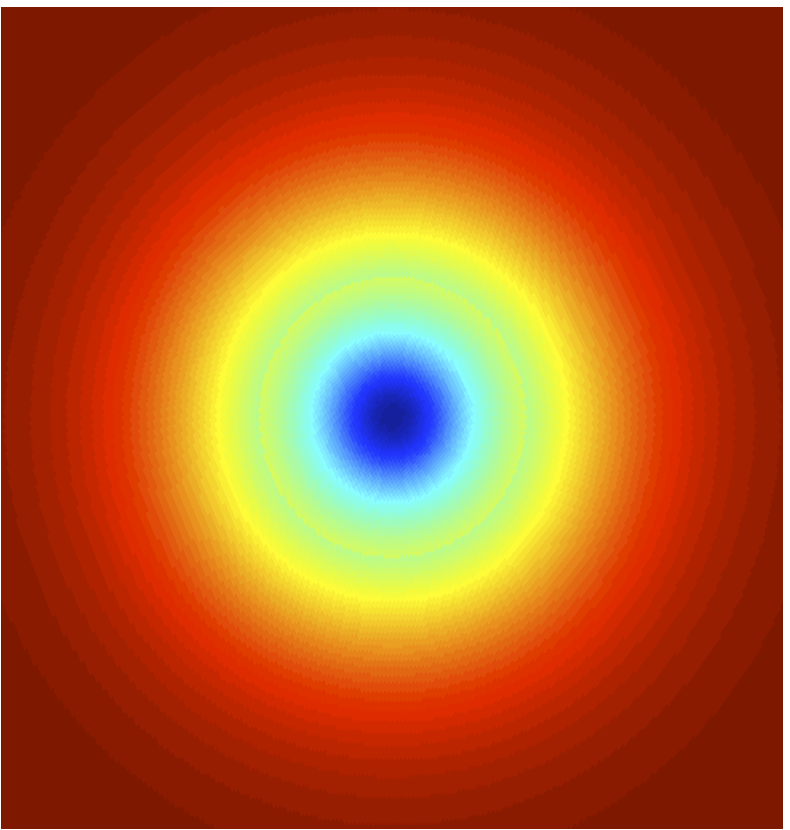}
\includegraphics[width=5.8cm,keepaspectratio]{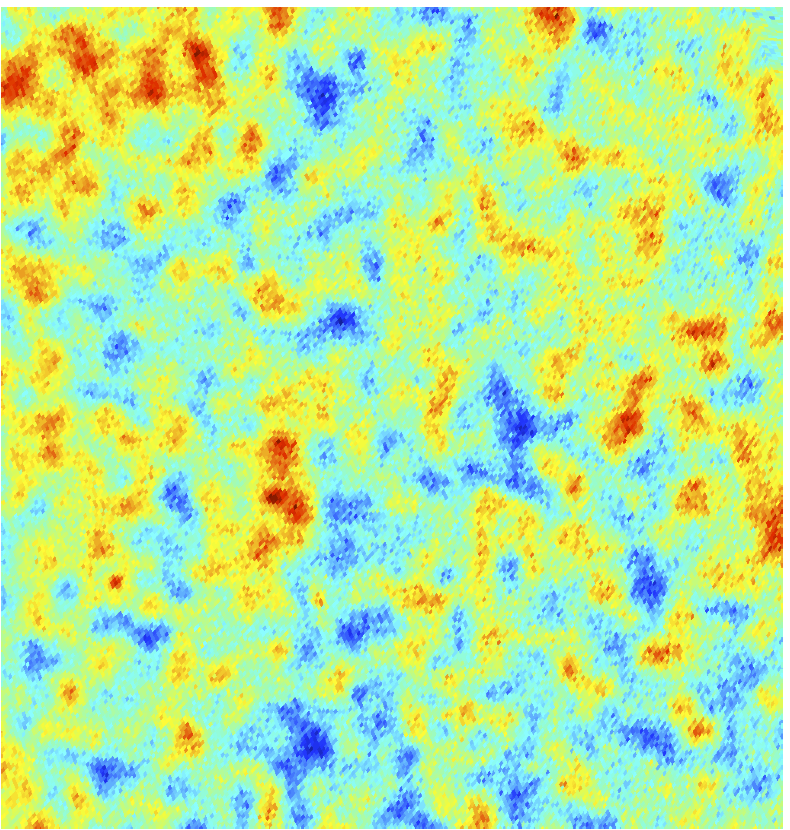}
\caption{~\label{fig:texture}Left panel, CMB temperature fluctuations obtained from WMAP data, centred in the position of the \csp. 
Middle panel, best-fit of a texture profile to the data, according to equation~\ref{eq:text_profile2}. Right panel: CMB map, at the position of the \csp, after the subtraction of the best-fit texture model.}
\end{center}
\end{figure*}
%%%%

The figure~\ref{fig:texture} shows the effect of correcting the QVW map from the texture emission.
In the left panel the region of the sky where the \cs was identified is shown. The middle panel presents the best-fit of the texture profile, according to the parameters previously mentioned. On the right panel, I present the resulting map after the subtraction of the estimated cosmic texture contribution. The \cs is noticeably reduced. More quantitative measurements were made by~\cite{cruz07b}. It was proved that, if cosmic textures (adequate to the parameters fixed by the \cs analysis) were added to isotropic and Gaussian simulations of CMB signals, as seen in the QVW WMAP map, then the kurtosis of the SMHW became compatible. Even more,~\cite{bridges08} showed that the WMAP map (corrected from the cosmic texture contribution in the location of the \csp) was not compatible with anisotropic patterns for non-standard expansions of the Universe (in particular, for Bianchi VII$_h$ models), as it was previously the case for the uncorrected data (e.g.~\cite{jaffe05,bridges07}).

Finally, let me remark that~\cite{cruz08} proposed a similar approach as the one described in this Section to study (attending to spatial templates) whether SZ and RS could provided more suitable hypotheses than the standard isotropic and Gaussian model. This Bayesian analysis indicated that neither of these hypotheses is favoured.

\subsection{Follow-up tests}
\label{subsec:foll}

The studies described in Section~\ref{sec:sources} and in the previous subsection, clearly indicates that, among all the realistic sources that could explain the anomalous nature of the \csp, only the cosmic texture hypothesis remains as a feasible option. The results obtained by~\cite{cruz07b}, and reviewed in the previous subsection, have to be understood as a (clear) indication that the cosmic texture is plausible.
However, before accepting it \emph{as the final explanation}, it should be confirmed by additional tests. In particular, if the cosmic texture hypothesis would be the right one, then there are some clear predictions that could be tested (at least in the near feature, once ongoing/upcoming experiments as SPT, ACT, QUIJOTE and ALMA are fully operative).

In this subsection I comment on the three most obvious follow-up tests that could help to discard, or accept, the cosmic texture. These follow-up tests are: the searching of more textures, the local polarization of the CMB, and the local CMB lensing. These foreseen effects were firstly pointed out by~\cite{cruz07b}.

\subsubsection{Looking for more textures}
\label{subsubsec:more}

If a texture were found in the location of the \csp, then, attending to cosmic texture models, there should be more cold and hot spots randomly distributed across the sky. In fact, the distribution of CMB spots caused by cosmic textures follows a scale-invariant law:
\begin{equation}
\label{eq:nc}
N_{\mathrm{sp}} (>\vartheta_c)= \frac{4\pi\nu\kappa^3}{3\vartheta^2_c},
\end{equation}
i.e., the number of cold/hot spots with a scale equal or larger than $\vartheta_c$, is inversely proportional to $\vartheta^2_c$ 
(see~\cite{spergel91} for the grounds and~\cite{cruz07b} for a derivation). The $\nu$ and $\kappa$ parameters are associated with the physics of the cosmic texture models, and their specific meaning is
out of the scope of this review. Let me remark here that, according to state-of-the-art simulations, these parameters are well consistent with
$\nu\approx2$ and $\kappa\approx0.1$ values, when $\vartheta_c$ is expressed in radians. Before continuing the discussion, let me stress that the number of expected cold/hot spots with a scale $\vartheta_c$ equal or larger than $5^\circ$ (i.e., as the \cs scale) is $\approx 1$. In other words, the fact that we only found a \cs of a scale similar or larger than the \cs size is fully
consistent with the cosmic texture scenario. This fact, of course, is an extra support for the texture hypothesis causing the \cs emission.

Let me come back to equation~\ref{eq:nc}. A straightforward calculation tells as that, if the cosmic texture hypothesis
is correct, then the CMB temperature fluctuations should contain $\approx 28$ spots with a scale $\vartheta_c \geq 1^\circ$, and $\approx 7$ spots with a scale $\vartheta_c \geq 2^\circ$.
Poisson errors can be safely assumed for these numbers. Therefore, the number of cold/hot
spots expected in the WMAP data are $1 \lesssim N_{\mathrm{sp}} (>2^\circ) \lesssim 13$ at the 95\% confidence level. Current work is
in progress to check this prediction, by using a fast cluster nesting sampling 
algorithm |\textsc{MULTINEST}~\cite{feroz09}| to explore the posterior probability ratio. There are
well founded hopes to find new textures. In particular, some non-Gaussian analyses as~\cite{vielva07a,pietrobon08,gurzadyan09} reported
some hot/cold spots (in addition to the \csp) as potentially anomalous.

\subsubsection{The polarization of the CMB}
\label{subsubsec:pol}
The effect of a collapsing texture on the passing by CMB photons is nothing but a secondary anisotropy of the CMB fluctuations, whose origin is merely gravitational. Hence, the effect of such gravitational phenomenon on the E-mode polarization is almost negligible (only vector modes would be affected, which are well below the scalar mode contribution).

Strictly speaking, this lack of polarization is not a unique signature produced by cosmic textures. As I said, any secondary anisotropy of gravitational origin would cause it.
However, these other effects (as huge voids) are quite implausible explanations (see Section~\ref{subsec:rs}). For that reason, this effect is a valid follow-up test to probe the texture hypothesis.

The procedure is simple: to compare the E-mode polarization in the position of a temperature spot (as large and extreme as the \cs is), under two different hypotheses, the \emph{null} or $H_0$ one (i.e., the temperature spot is caused by a Gaussian fluctuation) and the \emph{alternative} or $H_1$ option (i.e., the temperature spot is a secondary anisotropy caused by the collapse of an evolving texture). 
In fact, as proposed by~\cite{vielva10b}, the best discriminating measurement is the T-E correlation, rather than simply the E signal. This cross-correlation is expected to be close to zero for the $H_1$ hypothesis.
The approach suggested in this work was to estimate the correlation of the T and E profiles around the position of the spot temperature signal. This statistic was computed for many simulations according to the $H_0$ and $H_1$ hypotheses, and a hypothesis test was performed, via the definition of an optimal Fisher discriminant statistic (e.g.~\cite{cowan98}).

%%%%
\begin{figurehere}
\begin{center}
\includegraphics[width=8cm,keepaspectratio]{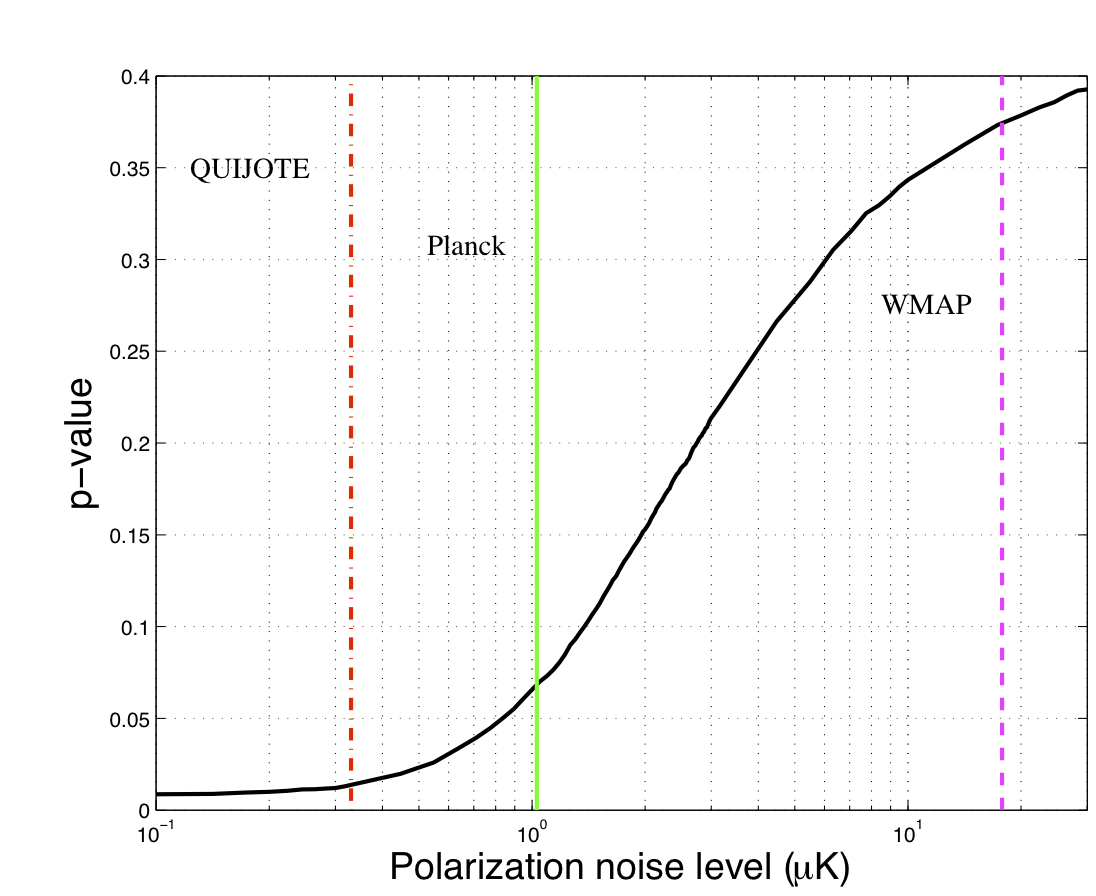}
\caption{~\label{fig:s2n}The curve indicates the probability (the p-value) for rejecting the texture hypothesis, attending to the T-E correlation, as a function of the instrumental noise sensitivity in polarization. From left to right, the vertical lines indicate the noise level associated with QUIJOTE, Planck, and WMAP experiments.}
\end{center}
\end{figurehere}
%%%%

The method was applied to probe the capabilities of current and upcoming CMB experiments for discriminating between the two hypotheses. In particular, the cases of WMAP, Planck~\cite{tauber10}, and QUIJOTE~\cite{rubinomartin10} were considered.

As the major conclusion, it can be established that the discrimination power of the T-E correlation is not very high. In fact, for an ideal noise-free experiment, at a power of the test of 0.5, the significance level is (up most) 0.8\%. The reason for this limitation is that the characteristic size of the \cs is $\approx 10^\circ$, which, roughly, corresponds to multipoles of $\ell \approx 40$. It happens that, at this multipole scales, the T-E angular cross-power spectrum is very close to zero, already for $H_0$ and, therefore, it is hard to discriminate it from $H_1$.

The results are graphically summarized in figure~\ref{fig:s2n}. I represent the significance level or p-value (for a power of the test of 0.5) as a function of the instrumental noise sensitivity for the E-mode polarization. The three experiments previously mentioned are indicated as vertical lines. Notice that the WMAP instrumental characteristics do not allow for any significant discrimination between hypotheses. With Planck, it would be possible to reach a modest 7\% significance level. Finally, the upcoming QUIJOTE experiment would allow for a 1.4\% detection. However, if the significance of the \cs obtained via the first non-Gaussianity wavelet analysis (i.e., 1.85\%, as mentioned in Section~\ref{sec:sig}) is also considered,~\cite{vielva10b} claims that a joint T and T-E significance detections of 0.025\% and 0.12\% could be imposed on the texture hypothesis, for the QUIJOTE and Planck experiments respectively.

\subsubsection{Gravitational lensing}
\label{subsubsec:lensing}
Besides the lack of polarization discussed in the previous subsection, the lensing of the CMB photons would be another foreseen effect caused by the gravitational field generated by the collapsing texture. This point was recently addressed by~\cite{das09}.
This work studied the capabilities of small-scale CMB experiments as ACT to detect a possible lensing effect occurring at the position of the \csp, caused by the gravitational field of a texture placed at $z=6$, and with a typical scale of $5^\circ$, associated with a symmetry-breaking energy scale of $\psi_0=4.5\times10^{15}$ GeV (i.e., comparable to the parameters determined by~\cite{cruz07b}). 

As for the polarization test, the power of the lensing analysis to probe the existence of the texture is relatively modest. For instance, detection is made at the 3$\sigma$ level after 1000 minutes of integration time. In other words, this test would require of a dedicated observational campaign.

\section{Conclusions}
\label{sec:final}

In this article I have presented a comprehensive overview of the \csp. 
Since its detection in 2003 by~\cite{vielva04},  this feature proved to be one of the most intriguing anomalies found in the WMAP data. 

The \cs was detected after performing a non-Gaussianity test on a cleaned CMB (obtained via a template fitting of the WMAP difference assemblies at Q-, V-, and W-bands).
The non-Gaussian analysis was performed by comparing the values obtained for the skewness and the kurtosis of the SMHW coefficients (at several scales) for such cleaned CMB map to the distribution expected from isotropic and Gaussian CMB realizations. This analysis indicated an excess of the kurtosis at SMHW scales $R$ of around 300 arcmin. Subsequent analysis of the SMHW coefficients, based on the area above/below a given threshold, the \emph{Higher Criticism}, and the maximum value, agreed in  detecting the non-Gaussian deviation occurring at the same scales, and confirmed the peculiar role played by the \csp.
Avoiding any possible \emph{a posteriori} choice of statistics, a conservative significance detection level of 1.85\% was placed by~\cite{cruz07a}.

The \cs was found to be highly isotropic~\cite{cruz06}, and the impact of possible systematics and  residual foreground contamination was discarded~\cite{vielva04,cruz05,cruz06,cruz07a}. Equivalently, some secondary anisotropies (potentially responsible for the \cs emission) as the Sunyaev-Zeldovich and the Rees-Sciama effects, that, in principle, could be accepted as valid solutions, were show to be unlikely explanations~\cite{cruz06,cruz08}.

A plausible explanation in terms of a cosmic defect was addressed by~\cite{cruz07b}. 
It would imply the presence of a collapsing cosmic texture at redshift $z\approx6$, with a typical scale of $\approx5^\circ$, corresponding to a symmetry-breaking energy scale of $\psi_0\approx 9\times10^{15}$. In addition, some follow-up tests were proposed to confirm or discard the cosmic texture hypothesis. In particular, one should expect to have more spots generated by evolving textures in CMB observations (e.g., $\approx7$ with scales $\vartheta \gtrsim 2^\circ$); a lack of E-mode polarization is predicted, as compared to the values associated with spots derived from a Gaussian CMB field; and, finally, a lensing of the CMB photons is potentially detectable with future small scale CMB experiments as ACT or SPT.

Summarizing, the study of the \cs has provided a wealth of information, and the upcoming high quality CMB data, as the one expected from Planck, guarantee that still more knowledge will come from the analysis of this very interesting feature.

\section*{Acknowledgements}
I thank R. Bel\'en Barreiro, Marcos Cruz and Enrique Mart\ii nez-Gonz\'alez for a careful reading of this article. The issue reviewed in this paper is the result of a very fruitful and exciting collaboration with them (and with other colleagues) during the last 7 years... and I presume it is not over yet, since new data is expected to come in the near future. I acknowledge partial financial support from the Spanish Ministerio de Ciencia e Innovaci{\'o}n project AYA2007-68058-C03-02. I also thank financial support from the \emph{Ram\'on y Cajal} programme.

\begin{bib}

\bibitem{abramo06} 
Abramo L.R., Bernui A., Ferreiro I.S., Villela T., Wuensche C.A., 2006, Phys. Rev. D, 74, 063506

\bibitem{aghanim99} 
Aghanim N., Forni O., 1999, A\&A, 347, 409

\bibitem{antoine98} 
Antoine J.-P., Vandergheynst P., 1998, Journal of Mathematical Physics, 39, 8, 3987

\bibitem{antoine04} 
Antoine J.-P., Murenzi R., Vandergheynst P., Ali S.T., 2004, \emph{Two-dimensional wavelets and their relatives}, Ed. The Cambridge
University Press, Cambridge, UK

\bibitem{ayaita10} 
Ayaita Y., Weber M., Wetterich C., 2010, Phys. Rev. D, 81, 023507

\bibitem{barreiro00}
Barreiro R.B., Hobson M.P., Lasenby A.N., Banda A.J., G\'orski K.M., Hinshaw G., 2000, MNRAS, 318, 475

\bibitem{barreiro01}
Barreiro R.B., Hobson M.P., 2001, MNRAS, 327, 813

\bibitem{bennett96}
Bennett C.L., et al., 1996, ApJ, 464, L1 

\bibitem{bevis04}
Bevis N., Hindmarsh M., Kunz M., 2004, Phys. Rev. D, 70, 043508 

\bibitem{bielewicz04}
Bielewicz P., G\'orski K.M., Banday A.J., 2004, MNRAS, 355, 1283 

\bibitem{bielewicz05}
Bielewicz P., Eriksen H.K., Banday A.J., G\'orski K.M., Lilje P.B.,
2005, ApJ, 635, 750

\bibitem{bremer10}
Bremer M.N., Silk J., Davies L.J.M., Lehnert M.D., 2010, MNRAS, 404, L69

\bibitem{bridges07}
Bridges M., McEwen J.D., Lasenby A.N., Hobson M.P., 2007, MNRAS, 377, 1473

\bibitem{bridges08}
Bridges M., McEwen J.D., Cruz M., Lasenby A.N., Hobson M.P., Vielva P., Mart\ii nez-Gonz\'alez E., 2008, MNRAS, 390, 1372

\bibitem{brough06}
Brough S. Forbes D.A., Kilborn V.A., Couch W., Colless M., 2006, MNRAS, 369, 1365

\bibitem{cao09}
Cao L., Fang, L.-Z., 2009, ApJ, 706, 1545 

\bibitem{carvalho09}
Carvalho P., Rocha G., Hobson M.P., 2009, MNRAS, 393, 681

\bibitem{cayon00}
Cay\'{o}n L., Sanz J.L., Barreiro R.B., Mar\ii nez-Gonz'alez E., Vielva P., Toffolatti L., Silk J., Diego J.M., Arg\"ueso F., 2000, MNRAS, 315, 757

\bibitem{cayon01}
Cay\'{o}n L., Sanz J.L., Mar\ii nez-Gonz\'alez E., Banday A.J., Arg\"ueso F., Gallegos J.E., G\'orski K.M., Hinshaw G., 2001, MNRAS, 326, 1243

\bibitem{cayon05}
Cay\'{o}n L., Jin J., Treaster A., 2005, MNRAS, 362, 826 

\bibitem{cayon10}
Cay\'{o}n L., 2010, MNRAS, in press 

\bibitem{colberg05}
Colberg J.M., Sheth R.K., Diaferio A., Liang G., Yoshida N., 2005, MNRAS, 360, 216

\bibitem{coles88}
Coles P., 1988, MNRAS, 234, 509

\bibitem{condon98} 
Condon J.J., Cotton W.D., Greisen E.W., Yin Q.F., Perley R.A., Taylor G.B., Broderick J.J., 1998, AJ, 115, 1693 

\bibitem{copi04} 
Copi C.J., Huterer D., Starkman G.D., 2004, Phys. Rev. D, 70, 043515 

\bibitem{copi06} 
Copi C.J., Huterer D., Schwarz D.J., Starkman G.D., 2006, MNRAS, 367, 79

\bibitem{cowan98} 
Cowan G., 1998, \emph{Statistical Data Analysis}, Oxford University Press

\bibitem{cruz05} 
Cruz M., Mart\ii nez-Gonz\'alez E., Vielva P., Cay\'on L., 2005, MNRAS,
356, 29

\bibitem{cruz06} 
Cruz M., Tucci M., Mart\ii nez-Gonz\'alez E., Vielva P., 2006, MNRAS,
369, 57 

\bibitem{cruz07a} 
Cruz M., Cay\'on L., Mart\ii nez-Gonz\'alez E., Vielva P., Jin J.,
2007a, ApJ, 655, 11

\bibitem{cruz07b} 
Cruz M., Turok N., Vielva P., Mart\ii nez-Gonz\'alez E., Hobson M.P.,
2007b, Science, 318, 1612 

\bibitem{cruz08}
Cruz M., Mart\ii nez-Gonz\'alez E., Vielva P., Diego J.M., Hobson M.P., Turok N., 
2008, MNRAS, 390, 913 

\bibitem{cruz10}
Cruz M., Vielva P., Mart\ii nez-Gonz\'alez E., Barreiro R.B.,  2010, (preprint arXiv:1005.1264)
 
\bibitem{curto09}
Curto A., Mart\ii nez-Gonz\'alez E., Barreiro R.B., 2009, ApJ, 706, 399

\bibitem{das09}
Das S., Spergel D.N., 2009, Phys. Rev. D, 79, 043007 

\bibitem{delabrouille09}
Delabrouille J., Cardoso J.-F., Le June M., Betoule M., Fay G., Gullaoux F., 2009,
A\&A, 493, 835

\bibitem{deOliveira04}
de Oliveira-Costa A., Tegmark M., Zaldarriaga M., Hamilton A., 2004,
Phys. Rev. D, 69, 063516

\bibitem{donoghue05}
Donoghue E.P., Donoghue J.F., 2005, Phys. Rev. D, 71, 043002 

\bibitem{donoho04}
Donoho D., Jin J., 2004, Ann. Statist., 2004, 32, 3, 962

\bibitem{durrer99}
Durrer R., Kunz M., Melchiorri A., 2004, Phys. Rev. D, 59, 123005

\bibitem{larson10}
Larson D, et al., 2010, (preprint arXiv:1001.4635)

\bibitem{eriksen04a}
Eriksen H.K., Hansen F.K., Banday A.J., G\'orski K.M., Lilje P.B.,
2004, ApJ, 605, 14

\bibitem{eriksen04b}
Eriksen H.K., Novikov D.I., Lilje P.B., Banday A.J., G\'orski K.M.,
2004, ApJ, 612, 64

\bibitem{fay08}
Fa\"y G., Guilloux F., Betoule M., Cardoso J.-F., Delabrouille J., Le June M., 2008, Phys. Rev. D,
78, 3013 

\bibitem{feroz09}
Feroz F., Hobson M.P., Bridges M., 2009, MNRAS, 398, 1601 

\bibitem{ferreira97}
Ferreira P.G., Magueijo J., Silk J., 1997, Phys. Rev. D, 56, 4592 

\bibitem{finkbeiner99}
Finkbeiner D.P., Davis M., Schlegel D.J., 1999, ApJ, 524, 867

\bibitem{finkbeiner03}
Finkbeiner D.P., 2003, ApJS, 146, 407

\bibitem{frommert10}
Frommert M., Ensslin T.A., 2010, MNRAS, 403, 1739

\bibitem{gold09}
Gold B., et al., 2009, ApJS, 180, 283 

\bibitem{gonzaleznuevo06}
Gonz\'alez-Nuevo J., Arg\"ueso F., L\'opez-Caniego M., Toffolatti L., Sanz J.L., Vielva P., Herranz D., 2006, MNRAS, 369, 1603

\bibitem{gorski05}
G\'orski K.M., Hivon E., Banday A.J., Wandelt B.D., Hansen F.K.,
Reinecke M., Bartelmann M., 2005, ApJ, 622, 759 

\bibitem{gott90}
Gott J.R., Park C., Juszkiwicz R., Bies W.E., Bennett D.P., Bouchet F.R., Stebbins A., 1990,
ApJ, 352, 1

\bibitem{granett09}
Granett B.R., Szapudi I., Neyrinck M.C., 2009, ApJ, 714, 825

\bibitem{gruppuso10}
Gruppuso A., Burigana C., 2010, JCAP, 8, 4

\bibitem{gurzadyan09}
Gurzadyan V.G., et al., 2009, A\&A, 497, 343

\bibitem{hammond09}
Hammond D.K., Wiaux Y., Vandergheynst P., 2009, MNRAS, 398, 1371

\bibitem{hansen04a}
Hansen F.K., Cabella P., Marinucci D., Vittorio N., 2004, ApJ, 607,
L67

\bibitem{hansen04b}
Hansen F.K., Banday A. J., G\'orski K.M., 2004, MNRAS, 354, 641 

\bibitem{hansen06}
Hansen F.K., Banday A. J., Eriksen H.K., G\'orski K.M., Lilje P.B., 2006, ApJ, 648, 748 

\bibitem{herranz02}
Herranz D., Sanz J.L.,Barreiro R.B., Mart\ii nez-Gonz\'alez E., 2002, ApJ, 580, 610

\bibitem{hobson99}
Hobson M.P., Jones A.W., Lasenby A.N., 1999, MNRAS, 309, 125

\bibitem{hoftuft09}
Hoftuft J., Eriksen H.K., Banday A.J., G\'orski K.M., Hansen F.K. Lilje P.B., 2009, ApJ,
699, 9856

\bibitem{inoue06}
Inoue K.T., Silk J., 2006, ApJ, 648, 23 

\bibitem{inoue07}
Inoue K.T., Silk J., 2007, ApJ, 664, 650 

\bibitem{jaffe05}
Jaffe T.R., Banday A.N., Eriksen H.K., G\'orski K.M., Hansen F.K., 2005, ApJ, 629, 1 

\bibitem{jeffreys61}
Jeffreys  H., 1961, \emph{Theory of Probability}, 3rd edn.  Oxford Univ. Press,  Oxford  

\bibitem{jonas98}
Jonas J., Baart E.E., Nicholson G.D., 1998, MNRAS, 297, 997

\bibitem{komatsu09}
Komatsu E., et al., 2009, ApJS, 180, 330

\bibitem{komatsu10}
Komatsu E., et al., 2010, (preprint, arXiv 1001.4538)

\bibitem{land05}
Land K., Magueijo J., 2005, MNRAS, 357, 994  

\bibitem{liddle00}
Liddle A., Lyth D. H., 2000, \emph{Cosmological inflation and large-scale structure},
Cambridge University Press

\bibitem{liu06}
Liu X., Zhang S.N., 2006, ApJ, 636, 1 

\bibitem{lopezcaniego06}
L\'opez-Caniego M., Herranz D., Gonz\'alez-Nuevo J., Sanz J.L., Barreiro R.B., Vielva P., 
Arg\"ueso F., Toffolatti L., 2006, MNRAS, 370, 2047

\bibitem{maisinger04}
Maisinger K., Hobson M.P., Lasenby A.N., 2004, MNRAS, 347, 339

\bibitem{martinezgonzalez90a}
Mart\ii nez-Gonz\'alez E., Sanz J.L., 1990, MNRAS, 247, 473

\bibitem{martinezgonzalez90b}
Mart\ii nez-Gonz\'alez E., Sanz J.L., Silk J., 1990, ApJ, 355, 5

\bibitem{martinezgonzalez02}
Mart\ii nez-Gonz\'alez E., Gallegos J.E., Arg\"ueso F., Cay\'on L., Sanz J.L., 2002, MNRAS, 336, 22

\bibitem{mcewen05}
McEwen J.D., Hobson M.P., Lasenby A.N., Mortlock D.J., 2005, MNRAS, 359, 1583 

\bibitem{mcewen07}
McEwen J.D., Vielva P., Wiaux Y., Barreiro R.B., Cay\'on L., Hobson M.P., Lasenby A.N., Mart\ii nez-Gonz\'alez E., Sanz J.L., 
2007, Journal Fourier Analyses and Applications, 13, 495 

\bibitem{mcewen08}
McEwen J.D., Wiaux Y., Hobson M.P., Vandergheynst P., Lasenby A.N., 2008, MNRAS, 384, 1289 

\bibitem{monteserin08}
Monteser\ii n C., Barreiro R.B., Vielva P., Mart\'inez-Gonz\'alez E.,
Hobson, M.P., Lasenby A.N., 2008, MNRAS, 387, 209 

\bibitem{moudden05}
Moudden Y., Abbrial P., Vielva P., Melin J.-B., Starck J.-L., Cardoso J.-F., Delabrouille J., Nguyen M.K., 2005,
Proc. Int. Conf. Physics in Signal and Image Processing (PSIP)

\bibitem{mukherjee00}
Mukherjee P., Hobson M.P., Lasenby A.N., 2000, MNRAS, 318, 1157 

\bibitem{mukherjee03}
Mukherjee P., Wang Y., 2003, ApJ, 599, 1 

\bibitem{mukherjee06}
Mukherjee P., Parkinson D., Liddle A,, 2006, ApJ, 638, L51 

\bibitem{mukherjee08}
Mukherjee P., Liddle A., 2008, MNRAS, 389, 209 

\bibitem{odgen97} 
Odgen R.T., 1997, \emph{Essential wavelets for statistical applications and data analysis}, Ed. Birkh\"auser, Boston, USA

\bibitem{paci10} 
Paci F., Gruppuso A., Finelli F., Cabella P., Da Rosa A., Mandolesi N., Natoli P., 2010, MNRAS, in press

\bibitem{pando98} 
Pando J., Valls-Gabaud D., Fang L.-Z., 1998, Phys. Rev. Lett., 81, 4568 

\bibitem{park04} 
Park C.-G., 2004, MNRAS, 349, 313 

\bibitem{pen94}
Pen U.-L., Spergel D.N., Turok N., 1994, Phys. Rev. D, 49, 692 

\bibitem{pietrobon06}
Pietrobon D., Balbi A., Marinucci D., 2006, Phys. Rev. D, 74, 043524 

\bibitem{pietrobon08}
Pietrobon D., Amblard A., Balbi A., Cabella P., Cooray A., Marinucci D., 2008, Phys. Rev. D, 78, 103504 

\bibitem{pietrobon10}
Pietrobon D., Cabella P., Balbi A., Crittenden R., de Gasperis G., Vittorio N., 2010, MNRAS, 402, 34

\bibitem{pires06}
Pires S., Juin J.B., Yvon D., Moudden Y., Anthoine S., Pierpaoli E., 2006, A\&A, 455, 741 

\bibitem{popa98} 
Popa L., 1998, New Astron., 3, 539

\bibitem{rath07}
R\"ath C., Schuecker P., Banday A.J., 2007, MNRAS, 380, 466 

\bibitem{rossmanith09}
Rossmanith G., R\"ath C., Banday A.J., Morfill G., 2009, MNRAS, in press 

\bibitem{rubinomartin10}
Rubi\~no-Mart\ii n J.A. et al., 2010, in \emph{Highlights of Spanish Astrophysics V}, 
Proceedings of the VIII Scientific Meeting of the 
Spanish Astronomical Society (SEA), Eds. J. Gorgas, L. J. Goicoechea, J. I. Gonz\'alez-Serrano, J. M. Diego 

\bibitem{rudjord09}
Rudjord O., Hansen F.K., Lan X., Liguori M., Marinucci D., Matarrese S., 2009,
ApJ, 701, 369

\bibitem{rudnick07}
Rudnick L., Brown S., Williams L.R., 2007, ApJ, 671, 40 

\bibitem{silvestri09}
Silvestri A., Trodden M., 2009, Phys. Rev. Lett., 103, 251301 

\bibitem{sanz99}
Sanz J.L., Arg\"ueso F., Mart\ii nez-Gonz\'alez E., Cay\'on L., Barreiro R.B., Toffolatti L., 1999, MNRAS, 309, 672

\bibitem{sanz06}
Sanz J.L., Herranz D., L\'opez-Caniego M., Arg\"ueso F., 2006
Proceedings of the 14th European Signal Processing Conference (EUSIPCO 2006), Eds. F. Gini and E.E. Kuruoglu

\bibitem{schmalzing98}
Schmalzing J., G\'orski K.M., 1998, MNRAS, 297, 355

\bibitem{schwarz04}
Schwarz D.J., Starkman G.D., Huterer D., Copi C.J., 2004, Phys. Rev.
Lett., 93, 221301

\bibitem{smith10}
Smith K.M., Huterer D., 2010, MNRAS, 403, 2 

\bibitem{spergel91}
Spergel D.N., Turok N., Press W.H., Ryden B.S., 1991, Phys. Rev. D, 43, 1038, 2 

\bibitem{sunyaev70}
Sunyaev R.A., Zeldovich Y.B., 1970, Ap\&SS, 7, 3

\bibitem{tauber10}
The Planck collaboration: Tauber J. et al., 2010, A\&A, in press 

\bibitem{tegmark03}
Tegmark M., de Oliveira-Costa A., Hamilton A., 2003, Phys. Rev. D, 68, 123523

\bibitem{tenorio99}
Tenorio L., Jaffe A.H., Hanany S., Lineweaver C.H., 1999, MNRAS, 310, 823 

\bibitem{turok89}
Turok N., 1989, Phys. Rev. Lett., 63, 2625

\bibitem{turok90}
Turok N., Spergel D.N., 1990, Phys. Rev. Lett., 64, 2736 

\bibitem{urrestilla08}
Urrestilla D., Bevis N., Hindmarsh M., Kunz M., Liddle A.R., 2008, JCAP, 07, 010 

\bibitem{vielva03}
Vielva P., Mar\ii nez-Gonz\'alez E., Gallegos J.E., Toffolatti L., Sanz J.L., 2003, MNRAS, 344, 89

\bibitem{vielva04}
Vielva P., Mart{\'\i}nez-Gonz\'alez E., Barreiro R.B., Sanz J.L.,
Cay\'on L., 2004, ApJ, 609, 22 

\bibitem{vielva06}
Vielva P., Mart{\'\i}nez-Gonz\'alez E., Tucci M., 2006, MNRAS, 365, 891 

\bibitem{vielva07a}
Vielva P., Wiaux Y., Mart\ii nez-Gonz\'alez E., Vandergheynst P., 2007, MNRAS, 381, 932 

\bibitem{vielva07b}
Vielva P., 2007, in \emph{Wavelets XII}. Edited by Van De Ville, Dimitri; Goyal, Vivek K.; Papadakis, Manos. Proceedings of the SPIE, 6701, 19,  

\bibitem{vielva09}
Vielva P., Sanz J.L., 2009, MNRAS, 397, 837

\bibitem{vielva10a}
Vielva P., Sanz J.L., 2010, MNRAS, 404, 895

\bibitem{vielva10b}
Vielva P., Mart{\'\i}nez-Gonz\'alez E., Cruz M., Barreiro R.B., Tucci M.,
2010, MNRAS in press (preprint arXiv:1001.4029)

\bibitem{vilenkin00}
Vilenkin A., Shellard E.P.S., 2000, \emph{Cosmic strings and other topological defects}, Cambridge University Press, Cambridge, UK

\bibitem{wiaux05}
Wiaux Y., Jacques L., Vandergheynst P.,
2005, ApJ, 632, 15 

\bibitem{wiaux06}
Wiaux Y., Vielva P., Mart\'inez-Gonz\'alez E., Vandergheynst P.,
2006, Phys. Rev. Lett., 96, 151303 

\bibitem{wiaux07}
Wiaux Y., McEwen J.D., Vielva P., 2007, Journal Fourier Analyses and Applications, 13, 477

\bibitem{wiaux08}
Wiaux Y., Vielva P., Barreiro R.B., Mart\'inez-Gonz\'alez E., Vandergheynst P., 2008, MNRAS, 385, 939 

\bibitem{zhang10}
Zhang R., Huterer D., 2010, Astropart. Phys., 33, 2, 69

\end{bib}

\end{body}
\end{document}